\documentclass[11pt]{article}
\pdfoutput=1
\usepackage{fullpage}
\usepackage{tensor}
\usepackage{xcolor}
\usepackage{cite}
\usepackage{graphicx}
\usepackage{amsmath}
\usepackage{amsthm}
\usepackage{amssymb}
\usepackage{subcaption}
\usepackage[toc,page]{appendix}
\usepackage[utf8x]{inputenc} 
\usepackage{framed}
\usepackage{mathrsfs}
\title{}
\date{}
\renewcommand{\vec}[1]{\mbox{\boldmath$ #1 $}}

\newcommand{\Indx}{\,\,\,}

\newtheorem{exmp}{Example}
\newtheorem*{principle}{Theorem}
\def\beq{\begin{equation}}
\def\eeq{\end{equation}}
\allowdisplaybreaks
\begin{document}
\bibliographystyle{utphys}

\newcommand\n[1]{\textcolor{red}{(#1)}} 
\newcommand{\diff}{\mathop{}\!\mathrm{d}}
\newcommand{\lb}{\left}
\newcommand{\rb}{\right}
\newcommand{\f}{\frac}
\newcommand{\pd}{\partial}
\newcommand{\tr}{\text{tr}}
\newcommand{\fdiff}{\mathcal{D}}
\newcommand{\im}{\text{im}}
\let\caron\v
\renewcommand{\v}{\mathbf}
\newcommand{\T}{\tensor}
\newcommand{\R}{\mathbb{R}}
\newcommand{\C}{\mathbb{C}}
\newcommand{\Z}{\mathbb{Z}}
\newcommand{\msbar}{\ensuremath{\overline{\text{MS}}}}
\newcommand{\DIS}{\ensuremath{\text{DIS}}}
\newcommand{\abar}{\ensuremath{\bar{\alpha}_S}}
\newcommand{\bb}{\ensuremath{\bar{\beta}_0}}
\newcommand{\rc}{\ensuremath{r_{\text{cut}}}}
\newcommand{\Nd}{\ensuremath{N_{\text{d.o.f.}}}}
\newcommand{\red}[1]{{\color{red} #1}}
\setlength{\parindent}{0pt}

\titlepage
\begin{flushright}
QMUL-PH-21-31\\
\end{flushright}

\vspace*{0.5cm}

\begin{center}
{\bf \Large The single copy of the gravitational holonomy}

\vspace*{1cm} \textsc{Rashid Alawadhi\footnote{r.alawadhi@qmul.ac.uk},
  David S. Berman\footnote{d.s.berman@qmul.ac.uk}, Chris
  D. White\footnote{christopher.white@qmul.ac.uk} and Sam
  Wikeley\footnote{s.wikeley@qmul.ac.uk}} \\

\vspace*{0.5cm} Centre for Research in String Theory, School of
Physics and Astronomy, \\
Queen Mary University of London, 327 Mile End
Road, London E1 4NS, UK\\

\end{center}

\vspace*{0.5cm}

\begin{abstract}
The double copy is a well-established relationship between gravity and gauge theories. 
It relates perturbative scattering amplitudes as well as classical solutions, and recently there has been mounting evidence that it also applies
to non-perturbative information. In this paper, we consider the {\it
  holonomy} properties of manifolds in gravity and prescribe a single copy of gravitational holonomy that
differs from the holonomy in gauge theory. We discuss specific cases and give examples where the single copy holonomy group is reduced. Our results may prove useful in extending the classical double copy. We also clarify previous
misconceptions in the literature regarding gravitational Wilson lines and holonomy.
\end{abstract}

\vspace*{0.5cm}

\section{Introduction}
\label{sec:intro}

The double copy~\cite{Bern:2010ue,Bern:2010yg} is by now a
well-established relationship between gauge theory and gravity, that
has been demonstrated for scattering amplitudes as well as classical
solutions. An important focus of ongoing research is to ascertain how
general the double copy is, and in particular whether it applies to
non-perturbative information. A successful understanding of the latter
may elucidate the underlying origin of the double copy, or reveal new
ways of thinking about different field theories, that make their
common structures manifest. \\

Previous attempts to study non-perturbative effects include analysing
strong coupling solutions of equations of
motion~\cite{White:2016jzc,DeSmet:2017rve,Bahjat-Abbas:2018vgo,Bahjat-Abbas:2020cyb,Berman:2020xvs},
examining exact algebras underlying the kinematic sectors of different
theories~\cite{Monteiro:2011pc,Borsten:2021hua,Chacon:2020fmr}, using
twistor
methods~\cite{White:2020sfn,Chacon:2021wbr,Farnsworth:2021wvs},
matching solution-generating transformations between different
theories~\cite{Alawadhi:2019urr,Huang:2019cja}, and studying whether
topological information (such as characteristic classes) can be
identified in gauge and gravity
solutions~\cite{Berman:2018hwd,Alfonsi:2020lub}. These studies suggest
that it is worthwhile to consider other global properties of gauge or
gravity solutions, and to ascertain whether or not they can be matched
according to a double copy prescription. In this paper, we study the
notion of {\it holonomy} which, loosely speaking, describes how a
vector is transformed after parallel transport around a closed
curve. The set of all such transformations forms a {\it holonomy
  group} and particular elements of the holonomy group are described
by path ordered integrals of the Christoffel connection along the
curve. In gauge theories, the equivalent concept is how the phase of a
charged particle transforms as it moves along a path. This is then
described by {\it Wilson line} operators, which involve integrating
the gauge field along a curve.\\

It has long been known that holonomy properties of gauge and gravity
theories are mathematically analogous. Gauge theories can be thought of in terms of
principal fibre bundles, where a base space (corresponding to
spacetime) is dressed by fibres acted on by the gauge group. The gauge
field itself is then associated with a connection on the fibre bundle. The description in gravity is
similar: one considers the tangent bundle obtained by dressing
spacetime with its tangent space at each point such that the tangent space is now the fibre. The connection on this
bundle corresponds to the connection in gravity. Thus, the holonomy in gauge and gravity theories share a common geometrical origin in that they both arise due to the fact that parallel transport in the base manifold induces transport in the fibre. It is therefore tempting to conclude
that the holonomy groups of gauge and gravity theories are directly
related by the double copy. As we will discuss in detail in this
paper, this assumption is false.\\

Attempts to explicitly relate holonomy properties and/or Wilson
lines in gauge and gravity theories have been made before. In
particular, refs.~\cite{Modanese:1991nh,Modanese:1993zh} studied the
holonomy properties of gravity solutions using a perturbative field
theory approach, based on Wilson line operators involving the
Christoffel symbol.\footnote{Similar gravitational Wilson lines have
  been used in lattice studies of quantum
  gravity~\cite{Hamber:2007sz,Hamber:2009uz}, and even date back to
  much earlier work~\cite{Mandelstam:1962us} that attempted to recast
  General Relativity in a manifestly coordinate-independent
  form.} The behaviour of these operators in perturbation theory was
found to be in striking contrast with the situation in gauge theory,
an observation that has arisen more recently in the study of
perturbative Wilson loops~\cite{Brandhuber:2008tf}. Given that the
double copy relates scattering amplitudes in perturbation theory, this
already suggests that the traditional holonomy operator in gravity is
not a double copy of its gauge theory counterpart. Indeed, there is a
second operator that one may write down in gravity, involving the
path-length of a particle traversing a curve, which has also been
called a Wilson
line~\cite{Hamber:1994jh,Brandhuber:2008tf,Naculich:2011ry,White:2011yy,Miller:2012an}. It
represents the phase experienced by a scalar particle that travels
around a closed loop (see also the earlier work of
ref.~\cite{Dowker:1967zz}), and in this sense is the correct physical
analogue of the holonomy operator in gauge theory, which arises in the
description of the Aharonov-Bohm effect. It has also been used in the
description of soft radiation~\cite{White:2011yy}, and high-energy
scattering~\cite{Melville:2013qca,Luna:2016idw}, in both cases
overlapping with results that can also be obtained via the double copy
of scattering
amplitudes~\cite{Oxburgh:2012zr,Vera:2012ds,Johansson:2013aca}. Recently,
ref.~\cite{Alfonsi:2020lub} argued that the single copy of this second
operator is indeed the (unique) gauge theory Wilson line, and that it can be
used to express non-trivial topological information of gauge and
gravity solutions in a common language (see also
refs.~\cite{Plefka:2018dpa} for a related study from a different point
of view).\\

The above discussion begs the following question: does the holonomy operator in gravity have a single copy?  The aim of this paper is to systematically
explore this question, and thus construct a square of four
operators, such that we have a pair of gauge theory operators which
are a meaningful single copy of the two gravity operators mentioned
above. Doing so will allow us to clear up some confusions in
the literature regarding the nature of Wilson lines in gravity, as well as provide yet more glimpses of potential
non-perturbative aspects of the double copy. \\

The structure of our paper is as follows. In section~\ref{sec:review},
we review relevant topics in the study of holonomy and Wilson
lines. In section~\ref{sec:singlecopy}, we discuss the single copy of
the Riemannian holonomy, as well as its relationship to an alternative
holonomy operator involving the spin connection, which will prove
useful for what follows. In section~\ref{sec:results}, we present
explicit results for the single copy operator, performing a detailed
comparison with its gravitational counterpart. Finally, we discuss our
results and conclude in section~\ref{sec:conclude}.

\section{Holonomy and Wilson lines}
\label{sec:review}

As discussed above, holonomy refers, in general, to the change in
certain mathematical objects as they are transported around a closed
loop. The particular description depends on which theory we are in, as
well as which quantities are being transported. In this section, we
review the relevant ideas that we will need for the rest of the paper.

\subsection{Riemannian holonomy}
\label{sec:riemann}

Given a (pseudo-)Riemannian manifold ${\cal M}$, one may consider the
tangent space $T_p({\cal M})$ of all vectors at a point $p$. The
tangent spaces $T_p({\cal M})$ and $T_q({\cal M})$ associated with
points $p,q\in{\cal M}$ will be different in general, such that one
must define a prescription for comparing vectors at different
points. As is well-known, the solution is to consider a curve
$\gamma(t)$ from $p$ to $q$, and to say that a given vector $V^\mu$
undergoes {\it parallel transport} along the curve if it satisfies the
equation
\begin{equation}
\frac{d}{dt}V^\mu+\Gamma^\mu_{\sigma\rho}\frac{dx^\rho}{dt}V^\sigma=0,
\label{parallel}
\end{equation}
where $\Gamma^\mu_{\sigma\rho}$ is the Christoffel symbol. Vectors
may be compared after they are parallel-transported to the same point (or
tangent space), and one may indeed solve eq.~(\ref{parallel}) to find
the total change in $V^\mu$ after it has been transported from $p$ to $q$:
\begin{equation}
V_q^\mu =[\Phi_\Gamma(\gamma)]^\mu {}_\sigma\, V_p^\sigma,\quad
[\Phi_\Gamma(\gamma)]^\mu {}_\sigma={\cal P}\exp\left[ -\int_\gamma dx^\rho
  \Gamma^{\mu}_{\rho\sigma} \right].
\label{Vchange}
\end{equation}
Here, in a slight abuse of notation, we have exponentiated the
Christoffel symbol considered as the matrix
$\tensor{[\Gamma_\rho]}{^\mu_\sigma}$. Furthermore, the path-ordering symbol
${\cal P}$ indicates that, in expanding the exponential, these
matrices are to be ordered according to increasing parameter value
$\tau$. A special case of eq.~(\ref{Vchange}) occurs if $q$ and $p$ are
taken to be the {\it same} point, on a closed curve $C$. The vectors
appearing on the left- and right-hand sides of eq.~(\ref{Vchange}) are
then living in the same tangent space, such that the change in $V^\mu$
upon being transported around $C$ is effected by the transformation
matrix
\begin{equation}
[\Phi_\Gamma(C)]^\mu {}_\sigma={\cal P}\exp\left[ -\oint_C dx^\rho
  \Gamma^{\mu}_{\rho\sigma} \right],
\label{PhiGammadef}
\end{equation}
which we will call the {\it Riemannian holonomy operator}. The set of
all such transformations forms the {\it Riemannian holonomy
  group}. For a generic $d$-dimensional Riemannian manifold, one
expects this group to be the maximal set of possible transformations
on vectors in the tangent space, namely O($d$). If the manifold is
orientable this group will reduce to SO($d$). Further reductions occur
in other special cases~\cite{BSMF_1955__83__279_0}, the details of
which have been of much study in the mathematical literature. For
example, the holonomy group reduces to $\text{SU}(d/2)$ for Calabi-Yau
manifolds, to $\text{Sp}(d/4)$ for hyper-K\"ahler manifolds, and even
to $\text{G}_2$ for exceptional seven dimensional manifolds.\\

For the benefit of a more mathematical audience, it is worthwhile to
briefly review how holonomy can be defined in more formal terms, which
is the subject of the following section.

\subsection{Mathematical formulation of holonomy}
\label{sec:maths}

The approach taken in the mathematics literature is to study the
holonomy of a given manifold by examining the parallel structures it
contains. More precisely, let $E$ be a vector bundle over a base
manifold $M$, and $\nabla$ a connection on $M$. This then defines
parallel transport as above. For any piece-wise connected curve
$\gamma:[a,b]\subset \mathbb{R} \rightarrow M$ an isomorphism of the
vector spaces in the fibre of $E$ is defined as
\begin{equation}
    \tau_{\gamma}: E_{\gamma(a)}\rightarrow E_{\gamma(b)}.
\end{equation}
To introduce the holonomy group $\text{Hol}(\nabla)$, we fix a point $p\in M$ and parallel transport a section of the bundle $X_p\in\Gamma(E_p)$ along all piece-wise connected \emph{loops} at $p$. This defines the holonomy group of the connection $\text{Hol}_p(\nabla)$. If we restrict ourselves to null-homotopic loops (loops that are contractible to a point) then we find the restricted holonomy group $\text{Hol}^0_p(\nabla)$. If the manifold is simply-connected then $\text{Hol}_p^0(\nabla)=\text{Hol}_p(\nabla)$, with the obvious fundamental group homomorphism $\pi_1(M)\rightarrow \text{Hol}_p(\nabla)/\text{Hol}^0_p(\nabla)$. Since the holonomy groups of a connected manifold at different points are isomorphic, we can talk about the holonomy group of the connection $\text{Hol}(\nabla)\subseteq \text{GL}(d,E)$. \\

A section $X\in \Gamma(E)$ is called parallel if it is constant with respect to the connection, i.e. $\nabla X = 0$. This is equivalent to saying that $X$ is invariant under the parallel transport map
\begin{equation}
    \tau_\gamma : X_{\gamma(a)} \rightarrow X_{\gamma(b)},
\end{equation}
for any piece-wise connected path $\gamma: [a,b] \rightarrow M$. Equivalently, we may write $\tau_{\gamma}(X_{\gamma(a)}) = X_{\gamma(b)}$, which is the non-coordinate basis expression of eq.~(\ref{Vchange}). Let us state the fundamental principle\cite{Galaev:2015mxf}:
\begin{principle}
\textit{There exists a one-to-one correspondence between parallel sections $X$ of the bundle $E$ and vectors $X_p$ in the fibre $E_p$ invariant under $\rm{Hol}_p(\nabla)$}.
\end{principle}
In other words, finding on a given (pseudo-)Riemannian manifold geometric objects whose covariant derivative vanishes is equivalent to finding the invariants of the holonomy group\cite{Besse:1987pua}. To elucidate the above statements, we now consider some examples of common Riemannian manifolds.

\begin{exmp}\upshape
On \textbf{orientable Riemannian manifolds} one can define the metric tensor $g$ as a section on the tensor product of the dual tangent bundle with itself $g\in\Gamma(TM^*\otimes TM^*)$, so that the metric tensor at a point $p\in M$ is a map $g_p: T_pM \otimes T_pM \rightarrow \mathbb{R} $. There is a unique torsion-less connection called the Levi-Civita connection such that $\nabla g = 0$. Therefore, $g$ is called parallel. To find $\text{Hol}_p(\nabla)$, denote the group of linear transformations preserving $g$ by $\text{O}(T_pM,g_p)$. Since $g$ is parallel, we find $\text{Hol}_p(\nabla)\subset \text{O}(T_pM,g_p)$. Furthermore, as $T_pM\cong \mathbb{R}^d$ we identify the holonomy group with a subgroup in ${\rm SO}(d)$.
\end{exmp}
\begin{exmp}\upshape
\textbf{K\"ahler Manifolds} are complex manifolds of dimension $d=2n$ and have a closed symplectic form $\omega=g(J,\,\cdot)\in\Lambda ^2(M_\mathbb{C})$, where $J$ is the almost complex structure and $\nabla\omega=0$. Now not only the metric tensor has vanishing exterior covariant; the symplectic form $\omega$, which is invariant under the symplectic group ${\rm Sp}(2n,\mathbb{R})$, also has vanishing covariant derivative. Going by our philosophy, the holonomy group must both preserve lengths \textit{and} the symplectic form. The sought after group is then ${\rm U}(n)$. This can be seen as the intersection
\begin{equation}
    U(n)=SO(2n)\cap Sp(2n,\mathbb{R}).
\end{equation}
\end{exmp}
\begin{exmp}\upshape
\textbf{Hyper-K\"ahler manifolds} of dimension $d=4n$ have three almost complex structures $(I,J,K)$ with vanishing covariant derivative and obey the quaternionic relations $I^2=J^2=K^2=Id$ and $IJ=JI=-K$. The subgroup of ${\rm SO}(4n)$ that preserves the quaternionic almost complex structure is $\text{Sp}(n)\subset \text{SO}(4n)$. Therefore $\text{Hol(g)}=\text{Sp}(n)$. An example of this is the holonomy group of the self-dual Taub-NUT metric with $\text{Hol}(\nabla)={\rm Sp}(1)\cong {\rm SU}(2)$.
\end{exmp}

\subsection{The spin connection holonomy}
\label{sec:spinconnection}

In eq.~\eqref{parallel} we have defined parallel transport explicitly
with the Christoffel connection. One may also use an alternative
formalism involving the {\it spin connection}, which would in any
case be necessary if one were to consider the parallel transport of
spinors. Let us now describe the role of the spin connection.  As is
well known (see e.g. ref.~\cite{Carroll:2004st} for a pedagogical
summary), at each point in spacetime, one may introduce a set of
orthonormal basis vectors $\hat{e}_{(a)}$, related to the usual
tangent space vector basis $e_{(\mu)}\equiv\partial_\mu$ by
\begin{equation}
e_{(\mu)}= \tensor{e}{^a_\mu} \hat{e}_{(a)},\quad
\hat{e}_{(a)}= \tensor{e}{_a^{\mu}} e_{(\mu)},
\label{vierbein}
\end{equation}
which defines the {\it (inverse) vielbein} $\tensor{e}{^a_\mu}$
($\tensor{e}{_a^\mu}$), satisfying
\begin{equation}
\tensor{e}{^a_\mu} \tensor{e}{^b_\nu} \eta_{ab}=g_{\mu\nu},\quad \tensor{e}{_a^\mu} \tensor{e}{_b^\nu}g_{\mu\nu}=\eta_{ab}, \quad
\tensor{e}{^a_\mu}\tensor{e}{_b^\mu} = \delta^a_b, \quad \tensor{e}{^a_\mu}\tensor{e}{_a^\nu} = \delta^{\nu}_{\mu},
\label{vierbein2}
\end{equation}
where $\eta_{ab}$ is the Minkowski metric of the flat tangent space.\footnote{We use a $(-,+,+,+)$ metric signature throughout.} Thus, the Roman indices ($a,b,...$) are raised and lowered with this flat metric, while the Greek indices ($\mu, \nu,...$) are raised and lowered using the curved metric $g_{\mu \nu}$. \\

The vielbein may now be used to relate the components of an arbitrary vector $V^\mu$ at a point $p$ to the components of a vector in the tangent space at $p$ via
\begin{equation}
V^a=\tensor{e}{^a_\mu}(p) V^\mu, \quad V^{\mu} = \tensor{e}{_a^{\mu}}(p)V^a.
\label{Vadef}
\end{equation}
From the vielbein we may calculate the spin connection
$\tensor{\omega}{^a_b}$ using {\it Cartan's first structure
  equation}. In differential form language, with the torsion set to zero, this is:
\begin{equation}
d e^a + \omega^a{}_b \wedge e^b =0 \, .
\end{equation}
We may also invert this to write the components of the spin connection in terms of the vielbein:
\begin{equation}
(\omega_{\mu})^{a b}=\frac{1}{2} e^{a\nu}\left(\partial_{\mu} \tensor{e}{^b_\nu}-\partial_{\nu} \tensor{e}{^b_\mu}\right)-\frac{1}{2} e^{b\nu}\left(\partial_{\mu} \tensor{e}{^a_\nu}-\partial_{\nu} \tensor{e}{^a_\mu}\right)-\frac{1}{2} e^{a\rho} e^{b\sigma}\tensor{e}{^c_\mu}\left(\partial_{\rho} e_{c\sigma}-\partial_{\sigma} e_{c\rho}\right) \, .
\label{spinvierbein}
\end{equation}

A vector in the tangent space will then satisfy the parallel transport equation as in
eq.~(\ref{Vchange}), but now with the spin connection:
\begin{equation}
\frac{d}{dt}V^a+\tensor{(\omega_{\mu})}{^a_b}\frac{dx^\mu}{dt}V^b=0 \, .
\label{spinparallel}
\end{equation}

The solution of eq.~(\ref{spinparallel}), by direct analogy with
eq.~(\ref{Vchange}), is 
\begin{equation}
V_q^a=[\Phi_\omega(\gamma)]^a{}_b V_p^b,\quad
[\Phi_\omega(\gamma)]^a{}_b={\cal P}\exp\left[-\int_\gamma dx^\mu 
(\omega_\mu)^a{}_b\right],
\label{Vachange}
\end{equation}
where ${\cal P}$ denotes path-ordering of the matrices
$\tensor{(\omega_{\mu})}{^a_b}$ along the worldline. Choosing $p$ and $q$ to
correspond to the same point on a closed curve $C$, one obtains the
holonomy of the spin connection:
\begin{equation}
[\Phi_\omega(C)]^a{}_b={\cal P}\exp\left[-\oint_C dx^\mu (\omega_\mu)^a{}_b\right].
\label{spinholonomy}
\end{equation}
This form of the holonomy is straightforwardly related to the Riemannian
holonomy operator of eq.~(\ref{PhiGammadef}). Upon transforming both
sides of eq.~(\ref{Vchange}) to the orthonormal basis and rearranging, one
obtains~\cite{Modanese:1993zh}
\begin{equation}
[\Phi_\omega(p,q)]^a{}_b=\tensor{e}{^a_\mu}(p)\,[\Phi_\Gamma(p,q)]^\mu{}_\nu\, \tensor{e}{_b^\nu} (q),
\label{Phirel}
\end{equation}
so that for a closed curve one has
\begin{equation}
[\Phi_\omega(C)]^a{}_b= \tensor{e}{^a_\mu}\, [\Phi_\Gamma(C)]^\mu{}_\nu\, \tensor{e}{_b^\nu},
\label{Phirel2}
\end{equation}
where the two vielbeins on the right-hand side are evaluated at the same point. The physical interpretation of this
relation is straightforward. The Riemannian holonomy operator
tells us how the components of a vector transform after the vector has been transported around $C$. The expression in terms of 
the spin connection does the same, but in the orthonormal basis. Then, the two holonomy operators are related by a similarity transformation, which is the content of eq.~(\ref{Phirel2}).\\

Thus, in discussing the holonomy of a given manifold, one is free to use either. For our later
purposes, it is convenient to rewrite eq.~(\ref{spinholonomy}) yet
further. Noting that the spin connection is valued in the Lie algebra
of the Lorentz group, we may introduce explicit Lorentz generators
$M^{ab}$ via
\begin{equation}
(\omega_\mu)^c{}_d=\frac{i}{2}(\omega_\mu)_{ab}(M^{ab})^c{}_d,
\label{omegaM}
\end{equation}
where the normalisation factor arises from the components of the generators in the spin-1 representation:
\begin{equation}
(M^{ab})^c{}_d = i( \eta^{ac} \delta^b_d - \eta^{bc} \delta^a_d).
\label{Mspin1}
\end{equation}
The operator of eq.~(\ref{spinholonomy}) is then
\begin{equation}
[\Phi_{\omega}(C)]^c{}_d= \tensor{{\cal P}\exp\left[
-\frac{i}{2}\oint_C dx^\mu\, (\omega_\mu)_{ab}\,M^{ab}
\right]}{^c_d} \, .
\label{spinholonomy2}
\end{equation}
Here it is worth noting that this form allows us to easily extend the notion of holonomy to spinors. One replaces the generators in the spin-1 representation with the generators in the spin-(1/2) representation constructed from the associated Clifford algebra. Concretely, instead of $M^{ab}$ one uses $(\Gamma^{ab})^\alpha{}_\beta$ and the holonomy group will be generically $\text{Spin}(d)$ valued. Doing this relies on lifting the tangent bundle of the manifold to a spin bundle, which requires the existence of a spin structure on the manifold. There may of course be global obstructions to doing this which are given by the second Stiefel-Whitney class. In what follows we will not worry about such spinor-valued holonomies, however it is certainly worth understanding how the double copy works in this case and how different representations in the tangent bundle are related to representations in the single copy. Having described how holonomy works in gravity, let us now consider
gauge theory.

\subsection{Holonomy in gauge theory}
\label{sec:gauge}

Consider a gauge theory defined on a spacetime manifold ${\cal M}$
with gauge group $G$. A field $\Psi^a$ transforming in a particular
representation of the gauge group can then be defined as a section of
a principal fibre bundle, where the gauge field itself is associated
with the connection. If we want to compare field values at different
points $p,q\in{\cal M}$, we must transform the gauge information
according to a suitable definition of parallel transport, leading to
an equation analogous to eq.~(\ref{Vchange}):
\begin{equation}
\Psi^a_q=[\Phi_A(\gamma)]^a{}_b\, \Psi^b_p,\quad 
[\Phi_A(\gamma)]^a{}_b= \tensor{{\cal P}
\exp\left[-g\int_\gamma dx^\mu {\bf A}_\mu\right]}{^a_b}.
\label{Phichange}
\end{equation}
Here ${\bf A}_\mu=A^a_\mu {\bf T}^a$ is the matrix-valued gauge field,
${\bf T}^a$ are the generators of the Lie algebra in the representation
appropriate to the field $\Psi^a$, and $g$ is the coupling. If
we again take $p$ and $q$ to be the same spacetime point lying on a
closed curve $C$, the change in $\Psi^a$ after transport around
the loop is given by
\begin{equation}
\Phi_A(C)={\cal P}\exp\left[-g\oint_C dx^\mu {\bf A}_\mu \right].
\label{Phidef}
\end{equation}
The set of all such transformations forms the holonomy group
associated with gauge theory solutions, which will be a subgroup of the gauge group $G$. \\

The operator appearing in eq.~(\ref{Phichange}) is known as a {\it
  Wilson line} in the gauge theory literature.\footnote{Note that we
  have chosen anti-Hermitian colour generators, such that there is no
  explicit factor of $i$ in the exponent of
  eq.~(\ref{Phichange}).} It transforms covariantly under gauge
transformations according to
\begin{equation}
\Phi_A(\gamma)\rightarrow {\bf U}_p\Phi_A(\gamma) {\bf U}_q^{-1},
\label{Phitrans}
\end{equation}
where ${\bf U}_p$ is an element of $G$ in the appropriate
representation, and represents a local gauge transformation at the
point $p$. Thus, Wilson lines are ubiquitous in the study of
scattering amplitudes, and typically crop up whenever some physical
behaviour has to be expressed in a gauge-covariant manner.  The
operator of eq.~(\ref{Phidef}) is known as a Wilson loop once the
trace is taken on the right-hand side. From eq.~(\ref{Phitrans}), this
renders the Wilson loop gauge-invariant. \\

Notably, the gravitational operator of eq.~(\ref{Vchange}) also
transforms similarly to eq.~(\ref{Phitrans}), but where the gauge
transformations are replaced by diffeomorphisms. That is, upon
changing coordinates according to $x^\alpha\rightarrow y^\alpha$, one
has~\cite{Modanese:1993zh}
\begin{equation}
[\Phi_\Gamma(\gamma)]^\mu{}_\sigma\rightarrow [\Lambda_p]^\mu{}_\alpha
\,[\Phi_\Gamma(\gamma')]^\alpha{}_\beta\,[\Lambda_q^{-1}]^\beta{}_\sigma,\quad
[\Lambda_p]^\lambda{}_\delta=\left(\frac{\partial y^\lambda}
{\partial x^\delta}\right)_p,
\label{PhiGammatrans} 
\end{equation}
where the path $\gamma$ is transformed according to the diffeomorphism
to $\gamma'$. This property, together with the fact that there is a
common geometric interpretation of the operators $\Phi_\Gamma(\gamma)$
and $\Phi_A(\gamma)$ in gravity and gauge theory respectively, has led
to $\Phi_\Gamma$ also being referred to as a Wilson line in the
gravity
literature~\cite{Modanese:1991nh,Modanese:1993zh,Hamber:2007sz,Hamber:2009uz}. However,
there is a more sensible candidate for this, as we discuss in the
following section.\\

\subsection{The gravitational Wilson line}
\label{sec:gravwilson}

In an abelian gauge theory, the Wilson line operator of
eq.~(\ref{Phidef}) has a useful physical interpretation, in that it
represents the phase change experienced by a charged particle as it
traverses a closed loop. The non-abelian version is a generalisation
of this, once the trace is taken to form a gauge-invariant
quantity. The analogous operator in gravity is easy to write down. In gravity the equivalent of the charge is the mass of the particle and so  the phase will only depend on the (Lorentz-invariant) path length of the closed curve multiplied by the particle mass. For an arbitrary
curve $\gamma$, one may then define the {\it gravitational Wilson
  line}~\cite{Hamber:1994jh} as follows (this is also discussed in the much earlier work of
ref.~\cite{Dowker:1967zz}):
\begin{equation}
\Phi_g(\gamma)=\exp\left[-im\int_\gamma d\tau 
\sqrt{-g_{\mu\nu}\dot{x}^{\mu}
\dot{x}^{\nu}}\right],
\label{Phigdef}
\end{equation}
where $m$ is the mass of the particle, $\tau$ its proper time, and $\dot{x}^{\mu} \equiv dx^\mu/d\tau$. Throughout this paper $\dot{x}^{\mu}$ will always denote differentiation with respect to the variable parameterising the curve. In
perturbation theory (as appropriate to the weak field limit), one may
introduce a graviton field $h_{\mu\nu}$ via
\begin{equation}
g_{\mu\nu}=\eta_{\mu\nu}+\kappa h_{\mu\nu},\quad \kappa=\sqrt{32\pi G_N}, 
\label{graviton}
\end{equation}
where $\eta_{\mu\nu}$ is the Minkowski metric, and $G_N$ Newton's
constant. Then the operator of eq.~(\ref{Phigdef}) simplifies, to first
non-trivial order in $\kappa$, to\footnote{We have ignored an overall
  multiplicative constant in eq.~(\ref{Phigdef2}), which will vanish
  in any vacuum expectation value of Wilson lines, once this is
  correctly normalised.}
\begin{equation}
\Phi_g(\gamma)=\exp\left[\frac{i\kappa}{2}\int_\gamma ds\, h_{\mu\nu}
\dot{x}^{\mu}\dot{x}^{\nu}\right],
\label{Phigdef2}
\end{equation}
where we have rescaled the integration variable to have mass dimension
$-2$. The square root in eq.~(\ref{Phigdef}) is cumbersome in general,
and one may further worry that the operator ceases to be defined for
massless particles. One may remove both problems by noting that the
exponent of eq.~(\ref{Phigdef}) contains the action for a point
particle. The latter can be replaced with the alternative
action\footnote{The use of this alternative point particle action in
a double copy context has been emphasised previously in
ref.~\cite{Plefka:2018dpa}.}
\begin{equation}
S_{\rm pp}= \frac{1}{2}\int d\tau \left[\frac{1}{e(\tau)}g_{\mu\nu}\dot{x}^{\mu}
\dot{x}^{\nu}-e(\tau)m^2 \right],
\label{Sdef}
\end{equation}
where $e(\tau)$ is an auxiliary parameter known as the {\it
  einbein}. Its field equation yields
\begin{equation}
\frac{\delta S}{\delta e}=-\frac{1}{2e^2}g_{\mu\nu}\dot{x}^{\mu}
\dot{x}^{\nu}-\frac{m^2}{2}=0,
\label{einbeineq}
\end{equation}
such that solving for $e(\tau)$ and substituting this into
eq.~(\ref{Sdef}) yields the original action of a point particle that
appears in eq.~(\ref{Phigdef}), in the massive case. The parameter
$e(\tau)$ plays the role of a ``metric'' on the worldline, and
transforms appropriately under reparametrizations. Choosing a value
for $e$ then amounts to fixing a gauge, and the choice $e=1$ in the
massless case immediately leads to the Wilson line of
eq.~(\ref{Phigdef2}). \\

It has recently been argued~\cite{Alfonsi:2020lub} that the operator
$\Phi_g(\gamma)$ is the double copy of the gauge theory Wilson line of
eq.~(\ref{Phichange}), which can be seen in a number of ways. It may be
related to scattering amplitudes, for example, by considering a
semi-infinite set of Wilson lines emanating from a common
point. Vacuum expectation values of such Wilson lines are known to
describe the infrared singularities of scattering amplitudes, where
the latter have been proven to formally double
copy~\cite{Oxburgh:2012zr}. That eq.~(\ref{Phigdef}) is the correct
Wilson line associated with IR singularities has been established in
refs.~\cite{Naculich:2011ry,White:2011yy}. Similar evidence comes from
the high energy (Regge) limit, where amplitudes are again known to
double-copy~\cite{Vera:2012ds,Johansson:2013aca}, and where there is
also a description in terms of the Wilson line operators of
eqs.~(\ref{Phichange})
and~(\ref{Phigdef2})~\cite{Korchemskaya:1994qp,Melville:2013qca,Luna:2016idw}. More directly, one may rewrite the operator in eq.~(\ref{Phichange}) as
\begin{equation}
\Phi(\gamma)={\cal P}
\exp\left[ig\tilde{\bf T}^a\int_\gamma ds  A^a_\mu\,\dot{x}^{\mu}\right],
\label{Phichange2}
\end{equation}
where we have temporarily adopted Hermitian colour generators defined
via $\tilde{\bf T}^a\equiv i {\bf T}^a$. If we consider a gauge field for which the double copy is known, then the gravitational Wilson line of
eq.~(\ref{Phigdef2}) is obtained by making the replacements
\begin{equation}
g\rightarrow\frac{\kappa}{2},\quad 
\tilde{\bf T}^a\rightarrow \dot{x}^{\mu},
\label{replacements}
\end{equation}
precisely mirroring the usual coupling and colour / kinematic
replacements associated with the BCJ double copy for
amplitudes. Further to the discussion in ref.~\cite{Alfonsi:2020lub},
it is interesting to note that the explicit double copy between the
gauge and gravity Wilson lines is manifest when using the alternative
point particle action of eq.~(\ref{Sdef}), for a particular choice of
einbein. This is perhaps not surprising: the original BCJ double copy
for amplitudes is known to be manifest only in certain {\it
  generalised gauges}, where conventional gauge transformations as
well as field redefinitions have potentially been applied. The same
property occurs also for classical solutions in general.  Thus, the
fact that a particular einbein is needed to get the double copy to
work -- itself a choice of gauge, as described above -- is entirely
consistent with previous instances of the double copy.\\

Aside from special kinematic limits of amplitudes, there are also
other situations in which $\Phi_g(\gamma)$ manifests itself as the
double copy of $\Phi_A(\gamma)$. It may be used, for example, to
quantify certain topological information that is the relevant
gravitational counterpart of that obtained in a gauge
theory~\cite{Alfonsi:2020lub}. Furthermore, a particular Wilson loop
involving $\Phi_A$ may be used to derive the Coulomb potential between
two static charges, such that replacing $\Phi_A$ with $\Phi_g$ instead
yields Newton's law of gravity~\cite{Hamber:1994jh}. That the two
potentials should indeed be related follows from the non-relativistic
limit of the classical double copy between the point charge and the
Schwarzschild black hole~\cite{Monteiro:2014cda}.\\

Given that the holonomy operator in gauge theory is related to the
Wilson line of eq.~(\ref{Phigdef2}) in gravity, it is therefore not
true that the holonomy operators in the two theories are related by
the double copy. The question then arises of whether one may find a
gauge theory single copy of the gravitational holonomy operator of
eq.~(\ref{PhiGammadef}). Indeed one can, as we explain in the
following section.

\section{The single copy of the gravitational holonomy}
\label{sec:singlecopy}

Above, we have seen that the holonomy operators in gauge and gravity
theory, whilst natural mathematical counterparts of each other, are
not physical counterparts in the sense of being related by the double/single copy. To find the correct single copy of the gravity result,
one must map the latter to a physical situation whose single copy is
already well-known. In the present case, the holonomy operator of
eq.~(\ref{spinholonomy2}) turns out to arise in the dynamics of
spinning particles, whose properties we review in the following
section.

\subsection{Relativistic spinning particles}
\label{sec:spinning}

In eq.~(\ref{Sdef}), we have seen the action for a spinless point
particle coupled to gravity. It is possible to generalise this to the
case in which a (possibly extended) object has an intrinsic angular
momentum (see ref.~\cite{Levi:2018nxp} for a modern pedagogical review, and also the classic works of refs.~\cite{Hanson:1974qy, Bailey:1975fe, Dixon:1970zza, Papapetrou:1951pa}). To this end, it is conventional to define a vierbein on the worldline, $\tensor{e}{^A_{\mu}}(\tau)$. The upper-case latin indices $(A,B,...)$ are those of a body-fixed frame; a frame fixed to the point-particle as it traverses the worldline. The vierbein $\tensor{e}{^A_{\mu}}$ therefore relates the body-fixed frame to the general coordinate frame. The angular velocity of the
object is then defined to be
\begin{equation}
\Omega_{\mu\nu}=e_{A\mu}\frac{D \tensor{e}{^A_\nu}}{D\tau},
\label{Omegadef}
\end{equation}
where 
\begin{equation}
\frac{D \tensor{e}{^A_\nu}}{D\tau} \equiv \dot{x}^\alpha D_\alpha \tensor{e}{^A_\nu}
=\dot{x}^\alpha\left(\partial_\alpha \tensor{e}{^A_\nu}-\Gamma^\lambda_{\alpha \nu}
\tensor{e}{^A_\lambda}\right)
\label{Dalpha} 
\end{equation}
is the spacetime covariant derivative of the body-fixed vierbein. That
eq.~(\ref{Omegadef}) satisfies the expected antisymmetry,
$\Omega_{\mu\nu}=-\Omega_{\nu\mu}$, follows from eq.~(\ref{vierbein2})
and the vanishing of the covariant derivative of the metric tensor
$g_{\mu\nu}$. The total action for the object can be written as
\begin{equation}
S_{\rm tot}=S_{\rm pp}+S_{\rm spin},
\label{Spp2}
\end{equation}
where $S_{\rm pp}$ is the spinless action of eq.~(\ref{Sdef}),
reflecting the fact that an extended object looks pointlike from a
sufficient distance. The correction term due to the spin
is\footnote{In writing eq.~(\ref{Sspin}), we have ignored additional
  gauge-fixing terms which are needed to eliminate residual arbitrary
  degrees of freedom in the precise definition of the spin
  tensor. Such terms will not matter for the arguments presented
  here.}
\begin{equation}
S_{\rm spin}=\int d\tau \left[\frac12 
\Omega_{\mu\nu}S^{\mu\nu}\right],
\label{Sspin} 
\end{equation}
where $S^{\mu\nu}$ is the {\it spin tensor} of the object, namely the
dynamical variable conjugate to the angular velocity. Physically this represents the intrinsic angular momentum of the object (in
either a classical or quantum setting). \\

In general, the vierbein $\tensor{e}{^a_\mu}$ that we choose for a given spacetime
will not correspond to the body-fixed vierbein $\tensor{e}{^A_\mu}$, and there will
therefore be a Lorentz transformation that relates the two:
\begin{equation}
{e^A}_\mu={\Lambda^A}_a {e^a}_\mu.
\label{etrans} 
\end{equation} 
The combination of terms appearing in eq.~(\ref{Sspin}) can then be
decomposed as~\cite{Vines:2016unv}
\begin{align}
S_{\mu\nu}\Omega^{\mu\nu}&=S_{\mu\nu}{\Lambda_A}^a {e_a}^\mu
\frac{D\Lambda^{Ab}{e_b}^\nu}{D\tau}\notag\\ &=S_{ab}\left({\Lambda_A}^a
\dot{\Lambda}^{Ab} - (\omega_\mu)^{ab}\dot{x}^\mu\right),
\label{omegadecompose}
\end{align}
where in the second line we have used eq.~(\ref{Dalpha}) together with
the known relation between the Christoffel symbol and spin connection
(see e.g.~\cite{Carroll:2004st})
\begin{equation}
\Gamma^\sigma_{\mu\nu}=\tensor{e}{_a^\sigma}\tensor{e}{^b_\nu}\tensor{(\omega_\mu)}{^a_b}
+\tensor{e}{_a^\sigma}\partial_\mu \tensor{e}{^a_\nu}.
\label{gamomega} 
\end{equation}

\subsection{The holonomy from a spinning particle}
\label{sec:holspin}

We now argue that the dynamics of spinning particles can be used to
construct a physical manifestation of the holonomy operator of
eq.~(\ref{spinholonomy2}). To this end, note that the two terms in
eq.~(\ref{omegadecompose}) have a straightforward physical
interpretation: the action of eq.~(\ref{Sspin}) dictates the dynamics
of the internal spin of the object under consideration, namely how the
body-fixed vierbein changes as one proceeds along the worldline. Put
another way, the action governs how a vector fixed to the moving
object will be modified, and there are clearly two distinct effects
causing such a vector to change: (i) the rotation of the object; (ii)
the fact that the body-fixed frame is changing due to the underlying
spacetime. These two effects are captured by the first and second terms
in the second line of eq.~(\ref{omegadecompose}) respectively, where
the first (rotation) term would be present even if the object were
moving in flat space.\\

In the previous section, we noted that the point particle action of
eq.~(\ref{Sdef}) could be used to form a Wilson line operator
representing the phase experienced by a particle as it traverses a
given contour. To do this, one forms the combination
\begin{equation}
e^{iS_{\rm pp}},
\label{Wilsonaction}
\end{equation}
and discards terms associated with flat space only i.e. that do not
involve the gravitational field. Such terms amount to an overall
multiplicative factor, that vanishes upon normalising vacuum
expectation values of Wilson lines. It is straightforward to repeat
this procedure for the action of eq.~(\ref{Spp2}), where the
corresponding Wilson line now represents the phase experienced by a
spinning particle. A given spin tensor will have the form
\begin{equation}
  S^{ab}(\tau)=Q^{ab}_{cd}(\tau) M^{cd}.
  \label{Sform}
\end{equation}
The quantity $Q^{ab}_{cd}$ denotes how much of each spin generator
is ``turned on'', and may in general depend on the parameter $\tau$
along the worldline. We wish to examine how all possible vectors are
transported around all possible loops. Thus, we may choose a spin
tensor such that
\begin{equation}
  Q^{ab}_{cd}=\frac12\left(\delta^{a}_{c}\delta^b_{d}
  -\delta^b_c\delta^a_d\right),
  \label{Qchoice}
\end{equation}
which physically amounts to a democratic assignment of unit spin along all axes. This discussion holds for a classical
particle. For a quantum particle in state $|\psi\rangle$, the spin
tensor will be given by a normalised expectation value
\begin{equation}
  S^{ab}=\frac{\langle\psi |Q^{ab}_{cd}\,M^{cd}
    |\psi\rangle}{\langle \psi|
\psi\rangle}.
\label{quantumspin}
\end{equation}
However, one may again make the choice of eq.~(\ref{Qchoice}), and for
concreteness we focus on a spin-1 particle with arbitrary
orientation. Furthermore, we will take our generalised Wilson line to
be matrix-valued in spin space, such that it describes how the spin
state of a test particle changes as it moves along its worldline. This
is directly analogous to how the gauge theory Wilson line of
eq.~(\ref{Phidef}) is matrix-valued in colour space, and in practical
terms amounts to the replacement
\begin{equation}
S^{ab}\rightarrow M^{ab}
\label{genspin}
\end{equation}
whilst defining the Wilson line according to the appropriate
generalisation of eq.~(\ref{Wilsonaction}). The result is
\begin{equation}
\Phi_g^{\rm spin}(\gamma)={\cal P}\exp\left[\frac{i\kappa}{2} \int_\gamma
ds \left(h_{\mu\nu}
\dot{x}^{\mu}\dot{x}^{\nu} - \dot{x}^{\mu}(\omega_\mu)_{cd}
M^{cd}\right)\right],
\label{Wilsonspin}
\end{equation}
where the path ordering is now necessary due to the matrix-valued
nature of the spin generators in the second term (note that there is also an
implicit identity matrix in spin space in the first term).  The first
term of eq.~(\ref{Wilsonspin}) represents how the mass of the test
particle couples to gravity, while the second term describes how its spin degrees of freedom couple to gravity. \\

As discussed above, a number of previous studies have attempted to
identify the operator of eq.~(\ref{PhiGammadef}) (and, by association,
the operator of eq.~(\ref{spinholonomy2})) as a gravitational Wilson
line, due to its mathematical similarity to the Wilson line in gauge
theory. However, the double copy tells us that
eq.~(\ref{spinholonomy2}) is not the gravitational counterpart to the
gauge theory Wilson line of eq.~(\ref{Phidef}). Rather, it is a
spin-dependent correction\footnote{Physically, the effects of the spin
term are suppressed by a power of the emitted graviton momentum, as we
discuss in section~\ref{sec:amplitudes}. So it is in this sense a
small correction to the spinless term in an appropriate kinematic
limit.} to the gravitational Wilson line of eq.~(\ref{Phigdef2}), and
represents the additional phase change that a particle experiences if
it happens to be spinning. It is interesting to note that a related
observation was made as early as the 1960s~\cite{Dowker:1967zz} (see
also ref.~\cite{Mandelstam:1962us}), predating the introduction of
Wilson lines!

\subsection{Single copy of the holonomy}
\label{sec:gaugetheory}

In the previous section, we have seen that a Wilson line constructed
from the action for a spinning particle coupled to gravity contains
the holonomy operator of eq.~(\ref{spinholonomy2}). This immediately
tells us how to take the single copy of the holonomy: we can simply
write down the action for a spinning particle coupled to a gauge
field, and use this to create a generalised Wilson line that contains
a spin correction to the phase. The relevant action for a spinning
particle coupled to a gauge field is (see e.g. ref.~\cite{Li:2018qap})
\begin{equation}
S_{\rm gauge}=\int d\tau \left[\frac{1}{2e(\tau)}\eta_{\mu\nu}
\dot{x}^{\mu}\dot{x}^{\nu}
-\frac{e(\tau)m^2}{2} +\frac12 \Omega_{\mu\nu} S^{\mu\nu}
+g c^a(\tau)
\left(\dot{x}^{\mu}{A}^a_\mu - \frac{e(\tau)}{2} F^a_{\mu\nu}S^{\mu\nu} 
\right)\right].
\label{gaugeaction}
\end{equation}
Here the first two terms are the usual point particle action of
eq.~(\ref{Sdef}) considered in Minkowski space. Furthermore,
$\Omega_{\mu\nu}$ is the flat space version of the angular velocity of
eq.~(\ref{Omegadef}), such that its contraction with the spin tensor
$S^{\mu\nu}$ matches the first term in the second line of
eq.~(\ref{omegadecompose}). There are then two terms involving the
gauge field, where $c^a(\tau)$ is a colour vector obtained by
evaluating the expectation value of the colour generator $\tilde{\bf T}^a$
at a given position on the worldline. The first of these terms gives
rise to the Wilson line operator of eq.~(\ref{Phidef}), once one
replaces the expectation value of the colour generator by the
generator itself. The remaining
term couples the field strength $F_{\mu\nu}^a$ to the spin tensor, and
thus represents the spin-dependent correction to the vacuum dynamics
of the spinning particle due to the presence of a gauge field. This is
the precise gauge theory analogue of the spin connection term in
eq.~(\ref{omegadecompose}), which amounts to the extra contribution to
the spin dynamics of the object arising from the gravitational
field.\\

Indeed, the double copy relationship between gauge theory and gravity
actions for spinning particles has been addressed in detail in
refs.~\cite{Goldberger:2017ogt,Li:2018qap}, which considered radiation
from such a particle interacting with a Yang-Mills field. The authors
calculated the effects of this radiation perturbatively, before
double-copying the results order-by-order in perturbation theory. The
resulting system is that of a spinning particle interacting with a
graviton, axion and dilaton, which is the usual spectrum arising from
the double copy of pure Yang-Mills theory. Thus, roughly speaking, the
final term in eq.~(\ref{gaugeaction}) double copies to multiple
operators, representing the coupling of the spin to the full field
spectrum in the gravity theory. This one-to-many nature of the double
copy does not affect our arguments here: given we are taking the {\it
  single copy}, which is many-to-one, we can unambiguously identify
the single copy of the graviton spin coupling as the final term
appearing in eq.~(\ref{gaugeaction}). Interestingly,
refs.~\cite{Goldberger:2017ogt,Li:2018qap} found that the spinning
particle actions in gauge theory and gravity were only strict physical
double copies of each other (in the sense that a conserved
energy-momentum tensor was found in the gravity theory) if a certain
fixed numerical coefficient was placed in front of the spin coupling
in the gauge theory. We shall ignore this complication here, given
that such a coefficient is irrelevant in elucidating the group of
transformations induced by the generalised Wilson line in the gauge
theory. Replacing the spin tensor according to eq.~(\ref{genspin}) as
before, as well as fixing the einbein $e=1$, this Wilson line is found
to be
\begin{equation}
\Phi_{\rm spin}(\gamma)={\cal P}\exp\left[ig\tilde{\bf T}^a\int_\gamma
ds \left( A_\mu^a \dot{x}^{\mu}-\frac{1}{2}F_{\mu\nu}^a M^{\mu\nu}
\right)\right]
\label{Phispin}
\end{equation}
Evaluating the spin correction over a closed curve $C$ then gives the
single copy of the gravitational holonomy:
\begin{equation}
\Phi_{F}(C)={\cal P}\exp\left[-\frac{ig}{2}\tilde{\bf T}^a\oint_C
ds F_{\mu\nu}^a M^{\mu\nu}\right].
\label{PhiFdef}
\end{equation}
As in the gravity theory, we can furnish the operator appearing in the
action of eq.~(\ref{gaugeaction}) with a physical interpretation,
where we may focus on the case of an abelian gauge theory for
simplicity. Identifying the electric and magnetic fields by 
\begin{equation}
F_{0i}=E_i,\quad F_{ij}=\epsilon_{ijk}B_k,
\label{EBfields}
\end{equation}
one finds
\begin{align}
F_{\mu\nu}S^{\mu\nu}&=F_{0i}S^{0i}+F_{ij}S^{ij}\notag\\
&=-\vec{E}\cdot \vec{d}-\vec{B}\cdot \vec{\mu},
\label{moments}
\end{align}
where we have defined
\begin{equation}
d_i=-S_{0i},\quad \mu_i=-\epsilon_{ijk}S^{jk}.
\label{moments2}
\end{equation}
Equation~(\ref{moments}) constitutes the electromagnetic coupling of
an electric dipole moment $\vec{d}$ and magnetic dipole moment
$\vec{\mu}$, and we may consider our test particle to have both of
these turned on in general.

\subsection{Relation to scattering amplitudes}
\label{sec:amplitudes}

In the previous section, we have used known results from the classical
double copy to identify the single copy of the gravitational
holonomy. We may also note, however, that our results can be linked to
known properties of scattering amplitudes. For example,
ref.~\cite{Arkani-Hamed:2019ymq} recently argued that the Kerr
(spinning) black hole and its single copy can be obtained from 3-point
amplitudes that are double-copies of each other. The implications of
this were explored further in
refs.~\cite{Emond:2020lwi,Guevara:2020xjx}. Similar conclusions have
been obtained independently in
refs.~\cite{Guevara:2018wpp,Guevara:2019fsj,Bautista:2019tdr}, which
also linked the double copy for spinning particles with the well-known
{\it (next-to-)soft theorems} for emission of low-momentum
radiation~\cite{Weinberg:1965nx,Gross:1968in,Low:1958sn,Burnett:1967km,DelDuca:1990gz,White:2011yy,Cachazo:2014fwa,Casali:2014xpa}. We
can see this directly from the above results as follows. Starting with
the Riemannian parallel transport operator of eq.~(\ref{Vchange}), we
can use the well-known relation between the Christoffel symbol and the
metric,
\begin{equation}
\Gamma^\mu_{\rho\sigma}=\frac12
g^{\mu\alpha}\left(\partial_\rho g_{\alpha\sigma }
+\partial_{\sigma}g_{\alpha\rho}-\partial_\alpha g_{\rho\sigma}\right),
\label{christoffelg}
\end{equation}
as well as the graviton definition of eq.~(\ref{graviton}), to obtain
\begin{equation}
\tensor{[\Phi_\Gamma(\gamma)]}{^\mu_\sigma}={\cal P}\exp\left[ -
\frac{\kappa}{2}\int_\gamma
  dx^\rho \left(\partial_\rho h^\mu_\sigma+\partial_\sigma h^\mu_\rho
  -\partial^\mu h_{\rho\sigma}+\ldots\right)\right],
\label{PhiGammaexp}
\end{equation}
where the ellipsis denotes terms of higher order in $\kappa$. Now
let us choose the case of a straight-line contour emanating from the
origin, as would be appropriate for a fast-moving particle with
momentum $p^\mu$ emerging from a scattering process:
\begin{equation}
x^\mu=s p^\mu,\quad 0\leq s < \infty.
\label{xparam}
\end{equation}
The first term in eq.~(\ref{PhiGammaexp}) is a total derivative, and
integrates to give a gauge-dependent artifact associated with the
endpoints of the contour. It will vanish for physical processes
e.g. in forming gauge-invariant amplitudes, or squaring amplitudes to
form a cross-section, which involves closing Wilson line contours to
make a closed loop. The remaining terms take the form
\begin{align}
-\frac{\kappa}{2} \int_0^\infty ds p^\rho \left(
\partial_\sigma h^\mu_\rho-\partial^\mu h_{\rho\sigma}\right)
&=\frac{i\kappa}{2}p^\rho \int\frac{d^dk}{(2\pi)^d}
\int_0^\infty ds \left(k_\sigma \tilde{h}^\mu_\rho 
-k^\mu\tilde{h}_{\rho\sigma}\right)e^{-isk\cdot p},
\label{NE1}
\end{align}
where we have introduced the Fourier components of the graviton field
via
\begin{equation}
h_{\mu\nu}=\int\frac{d^d k}{(2\pi)^d}\tilde{h}_{\mu\nu}(k)e^{-ik\cdot x}.
\label{Fourier}
\end{equation}
Carrying out the $s$ integral in eq.~(\ref{NE1})
yields\footnote{The upper limit of the $s$ integral in
  eq.~(\ref{NE1}) will vanish upon careful implementation of the
  Feynman $i\varepsilon$ prescription.}
\begin{equation}
\ln(\Phi_g)\sim \int\frac{d^d k}{(2\pi)^d}\tilde{h}_{\beta\rho}(k)
\left[\frac{\kappa}{2}\frac{p^\rho k_\alpha\tensor{(M^{\alpha\beta})}{^\mu_\sigma}}
{p\cdot k}\right],
\label{NE2}
\end{equation}
where we have written the second line in terms of the spin-1 Lorentz
generators of eq.~(\ref{Mspin1}). The square bracketed factor can be
recognised as the appropriate contribution to the next-to-soft theorem
for emission of a graviton~\cite{Cachazo:2014fwa}. Its appearance
in this context arises given that the spin-dependent coupling to the
worldline is suppressed by a single power of the momentum of emitted
radiation compared to the leading Wilson line operator of
eq.~(\ref{Phigdef}). Furthermore, one may perform an analogous calculation
for the gauge theory operator of eq.~(\ref{PhiFdef}), finding
\begin{equation}
\ln(\Phi_F)\sim \int\frac{d^d k}{(2\pi)^d}\tilde{A}^a_\mu(k)\left[
g\tilde{\bf T}^a\frac{k_\nu M^{\mu\nu}}{p\cdot k}\right],
\label{NE3}
\end{equation}
again in agreement with the appropriate next-to-soft
theorem~\cite{Casali:2014xpa}. The known double copy properties of
these results in the study of scattering amplitudes again corroborates
the fact that the single copy of the gravitational holonomy is the
operator of eq.~(\ref{PhiFdef}). It is also worth noting that these
generalised Wilson lines were derived before in the context of
next-to-soft physics, before the next-to-soft theorems were more
widely recognised~\cite{Laenen:2008gt,White:2011yy}. \\

Having presented a variety of arguments in favour of our single copy
holonomy operator, let us now note that further useful insights can be
obtained by focusing on a particular class of solutions, namely the
{\it Kerr-Schild} solutions in terms of which the first classical
double copy was formulated~\cite{Monteiro:2014cda}.

\subsection{Insights from Kerr-Schild solutions}

Kerr-Schild solutions of General Relativity (GR) are those for which the
metric assumes a particularly simple form, namely
\begin{equation}
    g_{\mu\nu}= \bar{g}_{\mu\nu}+\phi(x)k_\mu k_\nu.
\label{KSdef}
\end{equation}
Here $\bar{g}_{\mu\nu}$ is a background metric, which we will take to
be Minkowski throughout ($\bar{g}_{\mu\nu}=\eta_{\mu\nu}$), albeit not
necessarily in Cartesian coordinates. Furthermore, $\phi(x)$ is a
scalar field and $k_\mu$ a 4-vector field which is both null and
geodesic:
\begin{equation}
    \bar{g}^{\mu\nu}k_{\mu}k_{\nu}=g^{\mu\nu}k_\mu k_{\nu}=0,\quad k\cdot D k_{\mu}=0.
\label{KSconditions}
\end{equation}
Comparing with eq.~(\ref{graviton}), we see that Kerr-Schild solutions
have a graviton field given by
\begin{displaymath}
h_{\mu\nu}=\phi k_\mu k_\nu.
\end{displaymath}
This ansatz turns out to greatly simplify the Einstein equations,
which then have a linear dependence on the graviton only. This allows
for an infinite family of exact solutions to be obtained, which incudes
e.g. known black holes. Also, the ``factorised'' form of the graviton
(i.e. involving an outer product of a 4-vector with itself) allows a
single copy to be straightforwardly
obtained. Reference~\cite{Monteiro:2014cda} proved that for static
solutions, the gauge field
\begin{equation}
{\bf A}_\mu=A_\mu^a{\bf T}^a,\quad A_\mu^a=c^a \phi k_\mu,
\label{AKS}
\end{equation}
where $c^a$ is an arbitrary colour vector, solves the Yang-Mills
equations, which again simplify to a linear form. Consequently, the field strength tensor for the gauge field takes an abelian-like form:
\begin{equation}
    F^a_{\mu\nu}(x)=\pd_\mu A^a_{\nu}(x)-\pd_\nu A^a_{\mu}(x).
\label{fieldstrength}
\end{equation}
In an orthonormal basis, one has
\begin{equation}
    g_{\mu\nu}=\eta_{ab}\tensor{e}{^a_\mu}\tensor{e}{^b_\nu},
\label{gee}
\end{equation}
which in turn implies the following form for the Kerr-Schild vierbein:
\begin{equation}
        \tensor{e}{^a_{\mu}} = \tensor{\bar{e}}{^{\, a}_{\mu}} + \frac{1}{2}\phi k^ak_{\mu}, \quad
        \tensor{e}{_a^{\mu}} = \tensor{\bar{e}}{_{a}^{\mu}} - \frac{1}{2}\phi k_ak^{\mu}.
\label{eKS}
\end{equation}
Here $\tensor{\bar{e}}{^{\, a}_\mu}$ is the vierbein associated with the background metric in eq.~\eqref{KSdef}, which for our purposes is the Minkowski metric $\eta_{\mu\nu}$. 
As we review in appendix~\ref{AppSpin:A}, the spin connection
associated with this particular vierbein (subject to the additional
conditions of eq.~(\ref{KSconditions})) assumes the form
    \begin{equation}
        (\omega_{\mu})_{ab} = \pd_b e_{a\mu} - \pd_a e_{b\mu}.
\label{omegaKS}
    \end{equation}
Unlike the general expression of eq.~(\ref{spinvierbein}), this has
the pleasing property of being linear in the vierbein. Substituting
the results of eq.~(\ref{eKS}) (after lowering indices appropriately)
yields
    \begin{align}
        (\omega_{\mu})_{ab} &= \frac{1}{2}\left[\pd_b(\phi k_{\mu}k_a) - \pd_a(\phi k_{\mu}k_b)\right] \\
        &= \frac{1}{2}\left[\tensor{e}{_b^{\sigma}}\pd_{\sigma}(\phi k_{\mu}k_a) - \tensor{e}{_a^{\sigma}}\pd_{\sigma}(\phi k_{\mu}k_b)\right]. \label{scTemp1}
    \end{align}
Note that due to the null property of the Kerr-Schild vectors, $k^{\mu}k_{\mu} = 0$, conversion between coordinate and orthonormal bases is done simply with the background vierbein:
    \begin{equation}
        k_a = \tensor{e}{_a^{\mu}}k_{\mu} = \tensor{\bar{e}}{_{a}^{\mu}}k_{\mu} - \frac{1}{2}\phi k_ak^{\mu}k_{\mu}
        = \tensor{\bar{e}}{_{a}^{\mu}}k_{\mu}.
    \end{equation}
Thus, the spin connection in eq.~\eqref{scTemp1} can be written as
    \begin{equation}\label{scTemp2}
        (\omega_{\mu})_{ab} = \frac{1}{2}\left[\tensor{\bar{e}}{_{a}^{\nu}}\tensor{e}{_b^{\sigma}} - \tensor{\bar{e}}{_{b}^{\nu}}\tensor{e}{_a^{\sigma}}\right]\pd_{\sigma}(\phi k_{\mu}k_{\nu}).
    \end{equation}
If we now write the remaining vierbeins explicitly, a great deal of
simplification occurs. To see this, consider the first term in the
above expression:
    \begin{align}
        \tensor{\bar{e}}{_{a}^{\nu}}\tensor{e}{_b^{\sigma}}\pd_{\sigma}(\phi k_{\mu}k_{\nu}) 
        &= \tensor{\bar{e}}{_{a}^{\nu}}\left[\tensor{\bar{e}}{_{b}^{\sigma}} - \frac{1}{2}\phi k_bk^{\sigma}\right]\pd_{\sigma}(\phi k_{\mu}k_{\nu}) \\
        &= \tensor{\bar{e}}{_{a}^{\nu}}\tensor{\bar{e}}{_{b}^{\sigma}}\pd_{\sigma}(\phi k_{\mu}k_{\nu}) - \frac{1}{2}\phi k_ak_bk_{\mu}k^{\sigma}\pd_{\sigma}\phi.
    \end{align}
In the second equality, we have expanded $\pd_{\sigma}(\phi
k_{\mu}k_{\nu})$ using the product rule, from which two of the three
resulting terms vanish due to the geodesic condition
$k^{\sigma}\pd_{\sigma}k_{\mu}=0$. Performing the same procedure for
the second term in eq.~\eqref{scTemp2}, we find that the terms which
contain only a derivative of the scalar field $\phi$ cancel, such that
eq.~\eqref{scTemp2} is simply
    \begin{equation}
        (\omega_{\mu})_{ab} =
        \frac{1}{2}\left[\tensor{\bar{e}}{_{a}^{\nu}}\tensor{\bar{e}}{_{b}^{\sigma}} - \tensor{\bar{e}}{_{b}^{\nu}}\tensor{\bar{e}}{_{a}^{\sigma}}\right]\pd_{\sigma}(\phi k_{\mu}k_{\nu}).
    \end{equation}
Finally, if we expand the spin connection in terms of the Lorentz
generators, we obtain
    \begin{equation}
        \frac{i}{2}(\omega_{\mu})_{cd}M^{cd}
        = -\frac{i}{2}\pd_{\sigma}(\phi k_{\mu}k_{\nu})M^{\nu\sigma},
      \label{KSintegrand}
    \end{equation}
where we have identified the spin-1 Lorentz generators as
    \begin{equation}
        (M^{\nu\sigma})_{ab}
        = i\left[\tensor{\bar{e}}{_{a}^{\nu}}\tensor{\bar{e}}{_{b}^{\sigma}} - \tensor{\bar{e}}{_{b}^{\nu}}\tensor{\bar{e}}{_{a}^{\sigma}}\right].
    \end{equation}
Thus, we are left with a simple expression
in which the exponent appearing in the Kerr-Schild gravitational
holonomy operator is written directly in terms of the graviton:
    \begin{equation}\label{gravHol}
        \oint dx^{\mu}(\omega_{\mu})_{ab}M^{ab} = -\oint dx^{\mu}\pd_{\sigma}(h_{\mu\nu})M^{\nu\sigma}.
    \end{equation}
\newpage
The Kerr-Schild single copy of eq.~(\ref{AKS}) implies that we should
single copy eq.~(\ref{gravHol}) by replacing
\begin{equation}
  \dot{x}^\mu\rightarrow \tilde{{\bf T}}^a,\quad k_\mu\rightarrow c^a,
  \label{KSreplace}
\end{equation}
such that one obtains
    \begin{align}
        \oint dx^{\mu}(\omega_{\mu})_{ab}M^{ab} \to
        -\tilde{\bf T}^a \oint ds \, \pd_{\sigma}(\phi k_{\nu}c^a)M^{\nu\sigma}
        &= -\tilde{\bf T}^a
\oint ds \,\pd_{\sigma}(A^a_{\nu})M^{\nu\sigma} \label{SC}\\
        &= \frac{1}{2} \tilde{\bf T}^a\oint ds \, 
F^a_{\nu\sigma}M^{\nu\sigma}.
    \end{align}
This agrees with the conclusion reached above, namely that the single
copy of the gravitational holonomy is the operator of
eq.~(\ref{PhiFdef}).\\

By way of summarising the results of this section, we collect all of
the operators we have discussed in table~\ref{tab:operators}. The
gauge theory and gravity holonomies appear in the top-left and
bottom-right respectively, and the gravitational Wilson line appears
in the top-right. The operator in the bottom-left corner completes the
square, and has not been considered before as an analogue of the
gravity holonomy, due to its not having the appropriate differential
geometric definition. Indeed, its role is entirely different to the
gauge theory holonomy. Considering a gauge field via a connection on a
principal fibre bundle, the usual holonomy describes how vectors in
the internal colour space (associated with the fibres) are transported
as one moves along a curve. By contrast, the single copy of the
gravity holonomy instead describes how spacetime vectors are
transported, thus linking the gauge field with vectors living in the
tangent space of the base manifold. Our hope is that this operator
might also prove useful in classifying properties of different
Yang-Mills solutions, and we develop this notion in the following
section.
\renewcommand{\arraystretch}{2}
\begin{table}
\begin{center}
\begin{tabular}{c|c}
 Gauge Theory & Gravity\\
\hline
${\cal P}\exp\left[-g\oint_C dx^\mu {\bf A}_\mu \right]$
 & $\exp\left[\frac{i\kappa}{2}\oint_C ds \dot{x}^\mu \dot{x}^\nu
h_{\mu\nu}\right]$\\
 ${\cal P}\exp\left[\frac{g}{2}\oint_C ds {\bf F}_{\mu\nu} M^{\mu\nu}\right]$
& ${\cal P}\exp\left[-\frac{i\kappa}{2}\oint_C dx^\mu (\omega_\mu)_{ab}M^{ab}\right]$
\end{tabular}
\caption{Holonomy operators in gauge and gravity theories, and their
  single / double copies.}
\label{tab:operators}
\end{center}
\end{table}

\section{Results for the single copy holonomy operator}
\label{sec:results}

The gravitational holonomy is useful in that it allows us to classify
solutions of General Relativity (and arbitrary manifolds more
generally) into qualitatively different types. In general, one expects
the holonomy group of a given manifold to be the most general group
acting on vectors in the tangent space, namely $SO(d)$ for a
$d$-dimensional orientable spacetime in Euclidean signature, or
$SO(1,d-1)$ for Lorentzian signature. In some cases, however, the
holonomy group reduces to a subgroup, and the classification of
manifolds based on this idea has been widely studied (see
e.g. ref.~\cite{BSMF_1955__83__279_0} for the most well-known
incarnation). This suggests that our single copy holonomy operator
might have a similar purpose. Certainly, the operator of
eq.~(\ref{PhiFdef}) defines a group of transformations for a given
gauge theory solution. For ease of reference, we shall refer
to eq.~(\ref{PhiFdef}) as the SCH operator (short for
``single copy holonomy operator''), and the resulting group of
transformations as the SCH group. It may well be that the
SCH group reduces for certain gauge fields. There is also the
interesting possibility of taking gauge theory solutions that are
known single copies of particular gravity solutions, and asking
if the SCH and holonomy groups match up! We will investigate
this by considering particular solutions, of increasing complexity.

\subsection{The Schwarzschild black hole}
\label{sec:schwarzschild}

Arguably the simplest non-trivial gravity solution is the
Schwarzschild solution. It may be sourced by a point mass $M$ sitting
at the origin, and has a known Kerr-Schild form involving spherical
polar coordinates $(t,r,\theta,\varphi)$, where the functions entering
eq.~(\ref{KSdef}) may be chosen as
\begin{equation}
\phi(r)=\frac{M}{4\pi r},\quad k_\mu=(1,1,0,0).
\label{SWKS}
\end{equation}
We can then use eq.~(\ref{gravHol}) to ascertain the holonomy group,
for which we must choose a number of different closed contours, and
see what the various elements of the holonomy group are. Let us first
choose a circular orbit at constant time $t$, in the equatorial plane,
parametrised by
\begin{equation}
C:\quad x^\mu= (0,0,0,\varphi),\quad \varphi\in[0,2\pi).
\label{phiparam}
\end{equation}
The integral appearing in the holonomy operator then reduces to
\begin{equation}
\oint_C d\varphi \, \partial_\sigma (h_{\varphi \nu})M^{\nu\sigma}=0,
\label{intphi0}
\end{equation}
where we have used the fact that the Kerr-Schild graviton implied by
eq.~(\ref{SWKS}) has no non-zero $h_{\varphi \nu}$ components. The
element of the holonomy group associated with $C$ is thus the identity
element. Furthermore, spherical symmetry implies that similar constant
time loops that are tilted with respect to the equatorial plane will
also have a trivial holonomy.\\

\begin{figure}
\begin{center}
\scalebox{0.6}{\includegraphics{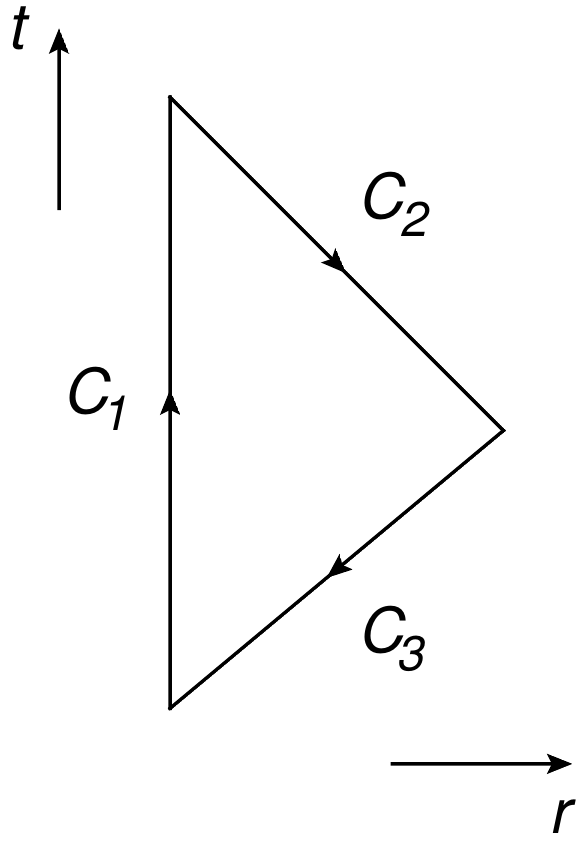}}
\caption{A loop consisting of three segments in the $(r,t)$ plane, where $C_2$ and $C_3$ are null lines.}
\label{fig:rtloop}
\end{center}
\end{figure}

To achieve a non-zero result, one may instead consider the curve shown
in figure~\ref{fig:rtloop}, consisting of three segments. 
The first segment $C_1$ is parallel to the time direction, and may be
parametrised by 
\begin{equation}
C_1:\quad x^\mu=(t,r_0,0,0),\quad 0\leq t\leq T,
\label{rtparam1}
\end{equation}
such that the curve is at a fixed radius $r=r_0$, and of total length
$T$. We have also chosen fixed values $\theta=\varphi=0$. From
eq.~(\ref{gravHol}), one finds
\begin{align}
\int_{C_1} dx^\mu \partial_\sigma (h_{\mu\nu})M^{\nu\sigma}=
\int_0^T dt \partial_r(h_{00})M^{0r}=
-\frac{MT}{4\pi r_0^2}M^{0r},
\label{C1int}
\end{align}
where we have noted the only non-zero contribution after contraction
of indices in the intermediate step. The remaining segments are
parametrised by
\begin{align}
C_2:\quad x^\mu&=(t,r_0+T-t,0,0),\quad T\geq t\geq T/2,\notag\\
C_3:\quad x^\mu&=\left(t,r_0+t,0,0\right),
\quad \frac{T}{2}\geq t\geq 0,
\label{rtparam2}
\end{align}
such that one finds
\begin{align}
\int_{C_2} dx^\mu \partial_\sigma (h_{\mu\nu})M^{\nu\sigma}=
\left.\int_T^{T/2}dt \partial_r(h_{00})M^{0r}\right|_{r=r_0+T-t} + 
\int_{r_0}^{r_0+T/2}
dr\partial_r(h_{r0})M^{0r};\notag\\
\int_{C_3} dx^\mu \partial_\sigma (h_{\mu\nu})M^{\nu\sigma}=
\left.\int_{T/2}^{0}dt \partial_r(h_{00})M^{0r}\right|_{r=r_0+t} + 
\int_{r_0+T/2}^{r_0}
dr\partial_r(h_{r0})M^{0r}.
\label{C23int}
\end{align}
One sees that the radial components cancel. Evaluating the remaining
integrals after using eq.~(\ref{SWKS}) gives
\begin{align}
\int_{C_2\cup C_3}dx^\mu \partial_\sigma (h_{\mu\nu})M^{\nu\sigma}
=\frac{M}{2\pi}\frac{T}{r_0(T+2r_0)}M^{0r},
\label{C23int2}
\end{align}
such that the total contribution from the entire loop is
\begin{align}
\oint dx^\mu \partial_\sigma (h_{\mu\nu})M^{\nu\sigma}=
\alpha M^{0r},\quad \alpha=-\frac{M}{4\pi}\frac{T^2}{r_0^2(T+2r_0)}.
\label{rtholonomy}
\end{align}
This constitutes an infinitesimal boost in the $(t,r)$ plane with
hyperbolic angle $\alpha$. To see this, we may recall the definition
of the boost generators $K_i$ and rotation generators $J_i$ in terms
of the $\{M^{\mu\nu}\}$:
\begin{equation}
K_i=M^{0i},\quad J_i=\frac12 \epsilon_{ijk}M^{jk},
\label{KJ}
\end{equation}
in terms of which the Lorentz algebra 
\begin{equation}
[M^{\mu\nu},M^{\rho\sigma}]=i\left(\eta^{\sigma\mu}M^{\rho\nu}
+\eta^{\nu\sigma}M^{\mu\rho}-\eta^{\rho\mu}M^{\sigma\nu}
-\eta^{\nu\rho}M^{\mu\sigma}\right)
\label{Lorentz}
\end{equation}
may be written as
\begin{equation}
[J_i,J_j]=i\epsilon_{ijk}J_k,\quad
[J_i,K_j]=i\epsilon_{ijk}K_k,\quad 
[K_i,K_j]=-i\epsilon_{ijk}J_k.
\label{KJalgebra}
\end{equation}
Equation~(\ref{rtholonomy}) is then clearly seen to contain the boost
generator $K_r$. If one considers loops with the same $r_0$ but
different fixed values of $\theta$ and $\varphi$, the full set of
boosts associated with arbitrary directions will be obtained. This in
turn implies that the holonomy group of the Schwarzschild spacetime is
SO$(1,d-1)$: from eq.~(\ref{KJalgebra}), one sees that the boosts do
not close upon themselves, such that exponentiating the boost
generators will produce transformations corresponding to combinations
of boosts and rotations in general. Note that our conclusions are in
qualitative agreement with those of e.g. ref.~\cite{Rothman:2000bz}
which also considered the holonomy in Schwarzschild spacetime. Our
explicit result for the boost angle differs due to having a loop with
a slightly different orientation, and also the use of Kerr-Schild
rather than conventional Schwarzschild coordinates.\\

Given this holonomy group, we may now consider the single copy, which
is well-known to be an abelian-like point charge in the gauge
theory~\cite{Monteiro:2014cda}. We may thus consider an abelian
exponent for the SCH operator:
\begin{equation}
\ln(\Phi_F)\rightarrow ig\oint_C ds F_{\mu\nu}M^{\mu\nu}.
\label{abelian}
\end{equation}
The only non-zero component of the field strength in this case
is
\begin{equation}
F_{0r}=\frac{Q}{4\pi r^2},
\label{pointchagre}
\end{equation}
where $Q$ is the charge. Plugging this into eq.~(\ref{abelian}), we
see immediately that an infinitesimal boost in the $(t,r)$ plane is
obtained, directly analogous to the Schwarzschild case. Thus, the
SCH group of the point charge is SO$(1,d-1)$. It is reasonable
to ponder at this point whether it is always the case that the
SCH and holonomy groups match up for gauge theory and gravity
solutions related by the double copy. That this is not the case will
be seen in the following example.

\subsection{Taub-NUT space}
\label{sec:NUT}

The Taub-NUT solution, first derived in refs.~\cite{Taub,NUT}, is a
non-asymptotically flat solution of GR, that has a rotational
character to the gravitational field at infinity. This is due to a
so-called {\it NUT charge} $N$, that is present in addition to a
Schwarzschild-like mass $M$. With a suitable choice of coordinates,
the metric may be written in Lorentzian signature as
    \begin{equation}\label{TNmetric}
        ds^2 = -A(r)\left[dt+B(\theta)d\phi\right]^2 + A^{-1}(r)dr^2 + C(r)\left[d\theta^2 + D(\theta)d\phi^2\right],
    \end{equation}
where for convenience we have defined 
    \begin{equation}\label{ABCD}
        A(r) = \frac{(r-r_+)(r-r_-)}{r^2 + N^2}, \quad
        B(\theta) = 2N\cos{\theta}, \quad
        C(r) = r^2+N^2, \quad
        D(\theta) = \sin^2{\theta},
    \end{equation}
and 
\begin{equation}
r_{\pm} = M \pm \sqrt{M^2 + N^2}. 
\label{rpm}
\end{equation}
The single copy of this solution was first considered in
ref.~\cite{Luna:2015paa}, and relied on the fact that coordinates
exist in which the Taub-NUT solution has a double Kerr-Schild
form~\cite{Chong:2004hw}. In the single copy, the mass $M$ maps to an
electric charge $Q$, as is familiar from the Schwarzschild case. The
NUT charge $N$, on the other hand, corresponds to a magnetic monopole
in the gauge theory, such that the single copy of the full Taub-NUT
solution is an electromagnetic dyon. This correspondence was
considered further in
refs.~\cite{Bahjat-Abbas:2020cyb,Alfonsi:2020lub}, which demonstrated how
magnetic monopoles in arbitrary non-abelian gauge theories can be
mapped to NUT charge in gravity. \\

For our present purposes, we want to examine the relationship (if
any) between the SCH group in gauge theory, and the holonomy
in gravity. Given that we have already seen this relationship for a
mass term $M$ in gravity, it is convenient to take this to zero, and
thus to consider the metric associated with a pure NUT charge $N$,
which maps to a magnetic monopole in gauge theory, whose magnetic
field may be written as
\begin{equation}
\vec{B}=\frac{\tilde{g}}{4\pi r^2}\hat{\vec{r}},
\label{Bmonopole}
\end{equation}
where $\tilde{g}$ is the magnetic charge, $r$ the spherical radius and
$\hat{\vec{r}}$ a unit vector in the radial direction. Note that we have
again chosen an abelian gauge group in the single copy for simplicity,
but the generalisation to a non-abelian context is
straightforward~\cite{Bahjat-Abbas:2020cyb,Alfonsi:2020lub}. From
eq.~(\ref{EBfields}), one sees that the only non-zero components of
the electromagnetic field strength are
\begin{equation}
F_{\theta\phi}=-F_{\phi\theta}=B_r=\frac{\tilde{g}}{4\pi r^2}.
\label{Fmonopole}
\end{equation}
Thus, the integral appearing in the SCH operator of
eq.~(\ref{PhiFdef}) reduces to
\begin{equation}
\oint_C ds \frac{\tilde{g}}{2\pi r} M^{\theta\phi},
\label{scholB}
\end{equation}
such that only the rotation generator in the $(\theta,\phi)$ plane is
turned on. This integral is indeed non-zero in general. Perhaps the
simplest case one may consider is a constant-time curve of radius
$r=r_0$ in the equatorial plane. All factors appearing in the
integrand in eq.~(\ref{scholB}) can then be taken outside the
integral, which then simply yields the length of the curve. Taking all
possible curves, the generator $M^{\theta\phi}$ will generate
rotations in all possible (Cartesian) directions, but not boosts. From
eq.~(\ref{KJalgebra}), one sees that the rotation algebra closes upon
itself. Thus, we straightforwardly obtain that the SCH group of
the magnetic monopole is SO(3) in four dimensions, and is therefore reduced
compared to the electric case of SO(1,3). \\

\begin{figure}
\begin{center}
\scalebox{0.6}{\includegraphics{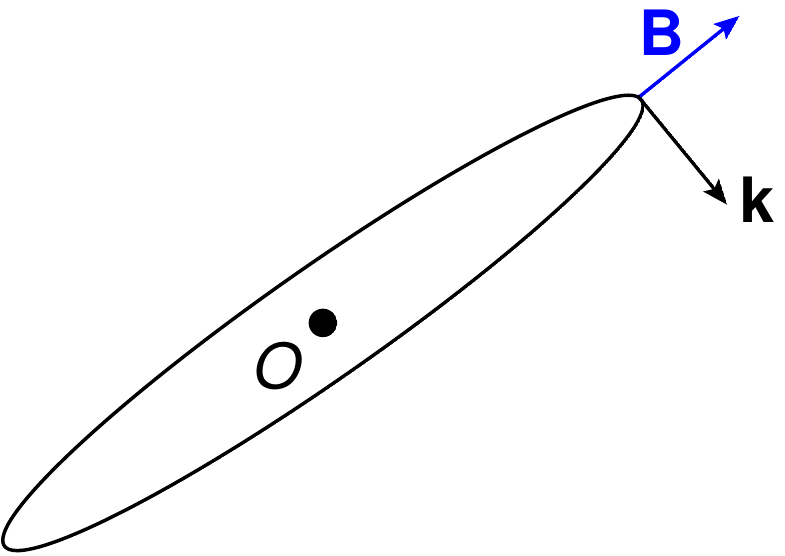}}
\caption{A closed spatial loop tilted with respect to the equatorial
  plane, where $O$ denotes the origin. A monopole magnetic field
  $\vec{B}$ generates a boost in the direction $\vec{k}$.}
\label{fig:boostloop}
\end{center}
\end{figure}

Let us now consider whether this matches up with the holonomy group of
the Taub-NUT solution in gravity. Given the metric of
eq.~(\ref{TNmetric}), we may choose the vierbein
    \begin{equation}\label{viel}
        e^0 = A^{\frac{1}{2}}(dt + Bd\phi), \quad
        e^1 = A^{-\frac{1}{2}}dr, \quad
        e^2 = C^{\frac{1}{2}}d\theta, \quad 
        e^3 = (CD)^{\frac{1}{2}} d\phi.
    \end{equation}
It will also be useful to have the inverse of these expressions:
    \begin{equation}\label{vielInverse}
        dt = A^{-\frac{1}{2}}e^0 - B (CD)^{-\frac{1}{2}}e^3, \quad
        dr = A^{\frac{1}{2}}e^1, \quad
        d\theta = C^{-\frac{1}{2}}e^2, \quad
        d\phi = (CD)^{-\frac{1}{2}}e^3.
    \end{equation}
The spin connection can be obtained from the torsion-free form of Cartan's structure equations,
    \begin{equation}\label{cartan}
        \tensor{\omega}{^a_b} \wedge e^b = -de^a,
    \end{equation}
along with the metric compatibility condition
    \begin{equation}\label{metricComb}
        \omega_{ab} = -\omega_{ba}.
    \end{equation}
Thus, we first calculate the exterior derivatives of the basis in eq.~\eqref{viel}, and use eq.~\eqref{vielInverse} to write the results in terms of the vielbein basis:
    \begin{align}
        de^0 &= (\pd_r A^{\frac{1}{2}})e^1 \wedge e^0 + (\pd_{\theta} B)C^{-1}\left(\frac{A}{D}\right)^{\frac{1}{2}}e^2 \wedge e^3, \\
        de^1 &= 0, \\
        de^2 &= (\pd_r C^{\frac{1}{2}}) \left(\frac{A}{C}\right)^{\frac{1}{2}}e^1 \wedge e^2, \\
        de^3 &= (\pd_r C^{\frac{1}{2}}) \left(\frac{A}{C}\right)^{\frac{1}{2}}e^1 \wedge e^3 + (\pd_{\theta} D^{\frac{1}{2}})\left(\frac{1}{CD}\right)^{\frac{1}{2}}e^2 \wedge e^3.
    \end{align}
These can now be used in the Cartan structure equations of eq.~\eqref{cartan} which, along with the metric compatibility condition in eq.~\eqref{metricComb}, yields the following non-zero components of the spin connection: 
    \begin{align}
        \tensor{\omega}{^0_1} &= \tensor{\omega}{^1_0} 
        = (\pd_r A^{\frac{1}{2}})A^{\frac{1}{2}}(dt + Bd\phi), \label{sc01} \\
        \tensor{\omega}{^0_2} &= \tensor{\omega}{^2_0} 
        = \frac{1}{2}(\pd_{\theta}B)\left(\frac{A}{C}\right)^{\frac{1}{2}}d\phi, \\
        \tensor{\omega}{^0_3} &= \tensor{\omega}{^3_0}
        = -\frac{1}{2}(\pd_{\theta}B)\left(\frac{A}{CD}\right)^{\frac{1}{2}}d\theta, \\
        \tensor{\omega}{^1_2} &= -\tensor{\omega}{^2_1}
        = -(\pd_rC^{\frac{1}{2}})A^{\frac{1}{2}}d\theta, \\
        \tensor{\omega}{^1_3} &= -\tensor{\omega}{^3_1}
        = -(\pd_rC^{\frac{1}{2}})(AD)^{\frac{1}{2}}d\phi \\
        \tensor{\omega}{^2_3} &= -\tensor{\omega}{^3_2}
        = -\frac{1}{2}(\pd_{\theta}B)\frac{A}{CD^{\frac{1}{2}}}(dt + Bd\phi)
        -(\pd_{\theta}D^{\frac{1}{2}})d\phi. \label{sc23}
    \end{align}
Now by substituting eqs.~\eqref{ABCD} and performing the derivatives we find for Taub-NUT:
    \begin{align}
        \tensor{\omega}{^0_1} &= \tensor{\omega}{^1_0}
        = \frac{M(r^2-N^2)+2N^2r}{(r^2+N^2)^2}\left[dt+2N\cos{\theta}d\phi\right], \label{omegaTN01} \\
        \tensor{\omega}{^0_2} &= \tensor{\omega}{^2_0}
        = -\frac{N\sin{\theta}}{r^2+N^2}\sqrt{(r-r_+)(r-r_-)}d\phi, \\
        \tensor{\omega}{^0_3} &= \tensor{\omega}{^3_0}
        = \frac{N}{r^2+N^2}\sqrt{(r-r_+)(r-r_-)}d\theta, \\
        \tensor{\omega}{^1_2} &= -\tensor{\omega}{^2_1}
        = -\frac{r}{r^2+N^2}\sqrt{(r-r_+)(r-r_-)}d\theta, \\
        \tensor{\omega}{^1_3} &= -\tensor{\omega}{^3_1}
        = -\frac{r\sin{\theta}}{r^2+N^2}\sqrt{(r-r_+)(r-r_-)}d\phi, \\
        \tensor{\omega}{^2_3} &= -\tensor{\omega}{^3_2}
        = \frac{N(r-r_+)(r-r_-)}{(r^2+N^2)^2}dt
        + \left[\frac{2N^2(r-r_+)(r-r_-)}{(r^2+N^2)^2} - 1\right]\cos{\theta}d\phi. \label{omegaTN23}
    \end{align}
For the single copy solution of a magnetic monopole above, we
considered a loop at constant time and radius in the equatorial plane
$\theta = \pi/2$. The integral in the holonomy operator is then simply
    \begin{equation}
        \oint dx^{\mu} (\omega_{\mu})_{ab} M^{ab} 
        = 2\oint d\phi [(\omega_{\phi})_{02} M^{02} + (\omega_{\phi})_{13} M^{13}]
        = 4\pi[(\omega_{\phi})_{02} M^{02} + (\omega_{\phi})_{13} M^{13}].
        \label{hTN}
    \end{equation}
This yields a boost in the 0-2 plane and a rotation in the 1-3 plane,
with coefficients $(\omega_{\phi})_{02}$ and $(\omega_{\phi})_{13}$
respectively, where
    \begin{align}
        (\omega_{\phi})_{02} &=
      \frac{N}{r^2+N^2}\sqrt{(r-r_+)(r-r_-)}, \label{holonomyNUT02}\\ (\omega_{\phi})_{13}
      &= -\frac{r}{r^2+N^2}\sqrt{(r-r_+)(r-r_-)}.
\label{holonomyNUT13}
    \end{align}
Note that the boost and rotation planes are mutually orthogonal, and
such a transformation is conventionally referred to as a {\it Lorentz
  four-screw}. Furthermore, our results are in agreement with those of
ref.~\cite{Bini:2002wd}, despite the different choice of vierbein
adopted by that reference. Equations~(\ref{holonomyNUT02},~\ref{holonomyNUT13}) still
correspond to the general Taub-NUT solution. We wish to consider the
double copy of the pure magnetic monopole, i.e. a pure NUT charge,
such that we may set $M\rightarrow0$ in eqs.~(\ref{holonomyNUT02},~\ref{holonomyNUT13}). The
integral of eq.~(\ref{hTN}) then becomes
    \begin{equation}
        \oint dx^{\mu} (\omega_{\mu})_{ab} M^{ab}=
        \frac{4\pi\sqrt{r^2-N^2}}{r^2+N^2}\left[
          N M^{02}-r M^{13}\right].
        \label{hTN2}
    \end{equation}
We thus see that the boost generator survives even in the case of a
pure NUT charge. By the arguments of the previous section, this will
potentially lead to the holonomy group SO(1,3), unless the effect of
the boost can be removed by performing a similarity transformation on
all group elements. However, upon considering other loops, boosts in
different Cartesian directions are generated. To see this, we may use
the fact that the metric for a pure NUT charge has a single
Kerr-Schild form, and thus we may use the expression of
eq.~(\ref{KSintegrand}) for the integrand of holonomy operator. The
coefficient of the boost generators, including the measure, is then
\begin{align}
dx^\mu \partial_\sigma(\phi k_\mu)M^{0\sigma}=
dx^\mu\left[\partial_\sigma(\phi k_\mu)-\partial_\mu(\phi k_\sigma)\right]
M^{0\sigma},
\label{hcalcKS}
\end{align}
where we have used the fact that $k_0=1$ for this
solution~\cite{lunaphd}, and in the second term we have introduced a total
derivative term that integrates to zero around a closed loop. From
eq.~(\ref{AKS}), we may recognise the expression in the closed brackets as the field strength tensor of a gauge field that is the single
copy of a Kerr-Schild graviton. One then finds

\begin{align}
dx^\mu \partial_\sigma(\phi k_\mu)M^{0\sigma}&=
dx^\mu F_{\sigma\mu}M^{0\sigma}\notag\\
&= dx_j \epsilon_{ijk} B_k K_i\notag\\
&=\vec{K}\cdot\left[d\vec{x}\times\vec{B}\right],
\label{hcalcKS2}
\end{align}
where we have used eqs.~(\ref{EBfields}, \ref{KJ}). The physical content of eq.~(\ref{hcalcKS2}) can be understood by considering a loop at constant time and
radius, that is tilted relative to the equatorial plane, as shown in
figure~\ref{fig:boostloop}.
The field of a magnetic monopole points radially outwards, whereas the
tangent 3-vector to the curve $d\vec{x}$ points into the page at the
point shown. This generates a boost in the direction $\vec{k}\propto
d\vec{x}\times\vec{B}$, which is easily seen to be in the increasing
$\theta$ direction. Thus, the conclusion reached above for the $\theta=\pi/2$ case, namely that there is a boost in the $(t,\theta)$
plane, turns out to be general for all such constant time loops. We
can therefore conclude that boosts in all Cartesian directions will be
turned on, such that the holonomy group of the pure NUT charge is
indeed SO(1,3), in contrast to the SCH group of the magnetic
monopole. Na\"{i}vely, one might have expected these groups to match
up, given that there are a well-known set of analogies between
electromagnetism and gravity known as {\it gravitomagnetism}. The pure
NUT charge is an extremal case of the Taub-NUT solution, that is most
like a magnetic monopole from a purely gravitational point of
view. However, the fact that its holonomy is sensitive to
gravitoelectric as well as gravitomagnetic effects is
well-documented~\cite{Bini:2002wd}, and arises from the fact that
spacelike hypersurfaces in the NUT solution have a nonzero extrinsic
curvature. This is not the case in the single copy gauge theory, which
lives in Minkowski space.

\subsection{Self-dual solutions}
\label{sec:self-dual}

The above examples probe different types of behaviour of the
SCH operator. For the Schwarzschild / point charge system, neither the
SCH nor the holonomy group reduce from their general form of
SO(1,3). For the NUT / monopole solutions, the SCH group does indeed
reduce, but the holonomy does not. In this section, we demonstrate
another possibility. Namely, that the SCH and holonomy groups
both reduce to mutually isomorphic subgroups of SO(1,3). We will
explicitly consider the case of {\it self-dual} solutions in gauge
theory and gravity. In gravity, one may decompose the Riemann tensor
into self-dual and anti-self-dual parts, given respectively by
\begin{equation}
R^\pm_{\mu\nu\rho\lambda}={(P^\pm)^{\alpha\beta}}_{\mu\nu} 
R_{\alpha\beta\rho\lambda},
\label{Rpmdef}
\end{equation}
where we have defined the projectors
\begin{align}
{(P^\pm)^{\alpha\beta}}_{\mu\nu} = \frac{1}{2}\left(\delta^{\alpha}_\mu\,\delta^\beta_\nu-
\delta^{\alpha}_\nu\,\delta^{\beta}_\mu\pm\sqrt{g}
{\epsilon^{\alpha\beta}}_{\mu\nu}\right),
\label{projectors}
\end{align}
with $g$ denoting the determinant of the metric.
Note also that we work in Euclidean signature throughout this section
only. We may use Stokes' theorem to rewrite the holonomy
operator of eq.~(\ref{spinholonomy2}) as~\cite{Dowker:1967zz}
\begin{align}
\Phi_\omega&={\cal P}\exp\left(-\frac{i}{2}\iint_\Sigma
d\Sigma^{\mu\nu}\,R_{cd\mu\nu}M^{cd}\right)\notag\\
&={\cal P}\exp\left(-\frac{i}{2}\iint_\Sigma
d\Sigma^{\mu\nu}\,(R^+_{\rho\lambda\mu\nu}
+R^-_{\rho\lambda\mu\nu})M^{\rho\lambda}\right).
\label{Stokes}
\end{align}
where $\Sigma$ is the area bounded by the curve $C$, with area element
$d\Sigma^{\mu\nu}$, and we have also converted vielbein indices to
coordinate frame indices in the second line. Note that $\Sigma$ is
simply connected, and therefore the holonomy group reduces to the
restricted holonomy group ${\rm Hol}^0(\nabla)$, which is only equal
to the full group $\text{Hol}(\nabla)$ when the fundamental homotopy
group $\pi_1$ is trivial. The second term is zero by definition for a
self-dual solution, in which case the holonomy operator becomes
\begin{align}
\Phi_\omega&={\cal P}\exp\left(-\frac{i}{2}\iint_\Sigma
d\Sigma^{\mu\nu}\, {(P^+)^{\alpha\beta}}_{\rho\lambda}R_{\alpha\beta\mu\nu}
M^{\rho\lambda}\right)\notag\\
&={\cal P}\exp\left(-\frac{i}{2}\iint_\Sigma
d\Sigma^{\mu\nu}\, R_{\alpha\beta\mu\nu}
(M^+)^{\alpha\beta}\right),
\label{SDhol2}
\end{align}
where we have defined two linearly independent sets of Lorentz
generators via
\begin{equation}
(M^\pm)^{\alpha\beta}={(P^\pm)^{\alpha\beta}}_{\rho\lambda}M^{\rho\lambda}.
\label{Mpm}
\end{equation}
This amounts to the known lie algebra isomorphism $\mathfrak{so}(4)\simeq \mathfrak{su}(2)\oplus \mathfrak{su}(2)$,
where each $\mathfrak{su}(2)$ subgroup corresponds to the generators
$(M^{\pm})^{\alpha\beta}$ respectively. It follows that for self-dual
solutions, the holonomy group reduces to ${\rm SU}(2)$, and
similar arguments can be applied to the case of anti-self-dual
solutions.\\

The single copy of a self-dual gravity solution is also self-dual in
the gauge theory~\cite{Monteiro:2013rya}. One may define the
(anti-)self-dual parts of the field strength as follows:
\begin{equation}
F^{\pm}_{\mu\nu}={(P^\pm)_{\mu\nu}}^{\alpha\beta}F_{\alpha\beta},
\label{FSD} 
\end{equation}
where one uses the projectors of eq.~(\ref{projectors}), but where now
the metric corresponds to that of Euclidean flat space (albeit
potentially in a curvilinear coordinate system, so that one must keep
the factor of $\sqrt{g}$). Self-dual solutions are defined by
$F^-_{\mu\nu}=0$, so that the SCH operator becomes
\begin{align}
\exp\left[ig\oint_C
ds F^+_{\mu\nu} M^{\mu\nu}\right]
&=\exp\left[ig\oint_C
ds {(P^+)_{\mu\nu}}^{\alpha\beta}F_{\alpha\beta} M^{\mu\nu}\right]\notag\\
&=\exp\left[ig\oint_C
ds F_{\alpha\beta}(M^+)^{\alpha\beta}
\right].
\label{PhiFSD}
\end{align}
Again only half of the generators are turned on, so that the SCH group
reduces to ${\rm SU}(2)$. The self-dual sector thus provides an interesting
example, in which the SCH and holonomy groups reduce, and are 
isomorphic. Furthermore, it is interesting to ponder whether the
arguments of this section generalise to manifolds of exotic holonomy
in higher dimensions, such as the well-known cases with holonomy
groups G$_2$ and Spin(7). It is not known how to explicitly single
copy such manifolds (see e.g. ref.~\cite{Alawadhi:2020jrv} for a
related discussion), but seeking Yang-Mills solutions with a suitable
SCH group might be a good place to start.

\section{Discussion}
\label{sec:conclude}

In this paper, we have considered the holonomy group in gravity, which
consists of the group of transformations acting on vectors that have
been parallel transported around a closed curve. The analogue of this in a (non)-abelian gauge theory is the
Wilson loop, which has a physical interpretation in terms of the phase
experienced by a scalar particle traversing a closed
contour. Attempts to match up the physical properties of the holonomy
and Wilson line -- or to interpret the holonomy operator itself as a
gravitational Wilson loop -- have been made
before~\cite{Modanese:1991nh,Modanese:1993zh,Hamber:2007sz,Hamber:2009uz,Brandhuber:2008tf},
with the conclusion that the gravitational holonomy should not be
thought as the being the correct physical analogue of the gauge theory
Wilson line. Indeed, a different gravitational Wilson line exists,
which corresponds to the phase experienced by a scalar
particle~\cite{Hamber:1994jh,Dowker:1967zz,Naculich:2011ry,White:2011yy}. This
begs the question of what the correct gauge theory analogue of the
gravitational holonomy is. To investigate this we have
used the {\it single copy}. We showed that the gravitational
holonomy arises naturally in the description of a spinning particle
interacting with a gravitatonal field. The single copy of this
situation is well-known to be a spinning particle interacting with a
gauge field. This allowed us to construct a generalised Wilson line
operator in the gauge theory, which gives the phase (non-diagonal in
spin space) experienced by a spin-1 test particle having an electric
and magnetic dipole moment. \\

Having found the single copy of the holonomy operator -- which we
dubbed the SCH operator -- we then commenced an
exploration of its properties.  We looked at certain special cases in which the
SCH group reduces, which includes the case of a pure magnetic
monopole, and also solutions which are self-dual. For the former, the
SCH group reduces even though the gravitational holonomy of the
monopole's double copy counterpart (a pure NUT charge) does not. \\

There are a number of avenues for further work. Firstly, one could
apply the SCH operator to different Yang-Mills solutions, and
see what general conclusions can be reached about their possible
SCH groups. It would also be interesting to look at how to
match the holonomy and its single copy in gauge and gravity theories more
generally, which might help in extending the classical double copy to
more complicated cases than are currently possible. Thirdly, it would
be nice if the SCH operator could shed light on
non-perturbative aspects of the double copy. In particular, we note
that the SCH operator is matrix-valued both in colour and spin
space. It thus rotates vectors both in the internal space associated
with the colour degrees of freedom, and also in the tangent space of
the manifold, which is associated with kinematic information. Might the single copy of the holonomy then have something to do with BCJ
duality~\cite{Bern:2008qj}, which links colour and kinematics in an
intriguing way?

\section*{Acknowledgments}
We thank Stefano De Angelis, Jung-Wook Kim, Adrian Padellaro, Malcolm
Perry and Rajath Radhakrishnan for many useful discussions.  We also
thank Jan Plefka for correspondence, some time ago, about the use of
the first-order point particle action in the double copy.  DSB is
happy to acknowledge the support of Pierre Andurand. This work has
been supported by the UK Science and Technology Facilities Council
(STFC) Consolidated Grant ST/P000754/1 ``String theory, gauge theory
and duality'', and by the European Union Horizon 2020 research and
innovation programme under the Marie Sk\l{}odowska-Curie grant
agreement No. 764850 ``SAGEX''. RA is supported by a student
scholarship from the Ministry of Education of the UAE.

\appendix

\section{Derivation of the Kerr-Schild spin connection}
\label{AppSpin:A}

In this appendix, we provide a derivation of the form of the spin
connection in Kerr-Schild coordinates, as reported in
eq.~(\ref{eKS}). The spin connection satisfies Cartan's first structure equation in the
absence of torsion,
\begin{equation}
    de^{a}+\omega^a_{\Indx c}\wedge e^c=0.
\end{equation}
In tensorial language this takes the form
\begin{equation}
    \pd_\mu \tensor{e}{^a_\nu} - \pd_\nu \tensor{e}{^a_\mu} + (\omega_\mu)^a_{\Indx \nu} - (\omega_\nu)^a_{\Indx \mu} = 0,
\end{equation}
were we have contracted the vierbein with the spin
connection. Multiplying by a factor of $\tensor{e}{_b^\mu}\tensor{e}{_a^\nu}$, one finds
\begin{equation}
     (\pd_b \tensor{e}{^a_\nu}) \tensor{e}{_a^\nu}
 - (\pd_b \tensor{e}{^a_\mu}) \tensor{e}{_b^\mu} + (\omega_b)^a_{\Indx c} - (\omega_c)^a_{\Indx b} = 0.
\end{equation}
Next, one can substitute the explicit forms of the Kerr-Schild
vierbein given in eq.~(\ref{eKS}), and use the null condition from
eq.~(\ref{KSconditions}), to obtain
\begin{equation}
     \pd_b \tensor{e}{^a_c} - \pd_c \tensor{e}{^a_b} + (\omega_b)^a_{\Indx c} - (\omega_c)^a_{\Indx b} - \frac{1}{4}\phi^2k^ak^{\mu} \left[ k_c \pd_b k_{\mu} -  k_b \pd_c k_{\mu} \right] = 0.
\end{equation}
Upon lowering the index $a$, one may cyclically permute the indices
$(a,b,c)$ and consider the combination $(a,b,c) - (b,c,a) - (c,a,b) = 0$, which yields 
\begin{equation}
    (\omega_{\mu})_{bc} = (\omega_a)_{bc}e^a_{\Indx \mu} = \left(\pd_c e_{ab} - \pd_b e_{ac} \right)e^a_{\Indx \mu} + \frac{1}{4}\phi^2k_{\mu}k^{\nu} \left( k_c\pd_b k_\nu - k_b \pd_c k_\nu  \right),
\end{equation}
where we have also multiplied the entire equation by $e^a_{\Indx \mu}$
to turn $a$ into a spacetime index. One may again use eqs.~(\ref{eKS},
\ref{KSconditions}) for the vierbein in the first term, after which
cancellations occur, leading to
\begin{equation}
    (\omega_\mu)_{ab} = \pd_b e_{a\mu} - \pd_a e_{b\mu}.
\end{equation}
This agrees with a similar result in ref.~\cite{Chakrabarti:1999mb}, and can also be obtained by plugging the Kerr-Schild vierbein of eq.~\eqref{eKS} into eq.~\eqref{spinvierbein}.


\bibliography{refs}
\end{document}